\documentclass[a4paper,final,onecolumn]{article}
\usepackage{geometry}
\pdfoutput=1

\usepackage[utf8]{inputenc}
\usepackage[english]{babel}
\usepackage{lineno}
\usepackage{amsmath}
\usepackage{amssymb}
\usepackage{mathtools}
\usepackage{miller}
\usepackage{graphicx}
\usepackage{bm}

\usepackage[binary-units=true]{siunitx}

\usepackage{xcolor}
\usepackage{authblk}
\usepackage[font=small,labelfont=bf]{caption}
 
\usepackage{csquotes}
\usepackage[backend=bibtex,style=numeric-comp,sorting=none]{biblatex}
\newcommand{\python}{Python}
\newcommand{\jupyter}{Jupyter}
\newcommand{\matlab}{MATLAB}
\newcommand{\mtex}{MTEX Toolbox}
\newcommand{\paraprobe}{PARAPROBE}

\newcommand{\imkl}{Intel Math Kernel Library}

\newcommand{\fftw}{The Fastest Fourier Transform in the West (FFTW)}
\newcommand{\cufft}{cuFFT}
\newcommand{\ivasfour}{IVAS}

\newcommand{\ashapes}{$\alpha$-shapes}

\newcommand{\bunge}[3]{($\varphi_1=\SI[mode=math]{#1}{\degree}, \Phi=\SI[mode=math]{#2}{\degree},\varphi_2=\SI[mode=math]{#3}{\degree}$)}

\newcommand{\swisswatch}[2]{$\SI[mode=math]{#1}{\hour}$ and $\SI[mode=math]{#2}{\minute}$}

\newcommand{\althreesc}{$\textnormal{Al}_{\textnormal{3}}\textnormal{Sc}$}
\newcommand{\althreeli}{$\textnormal{Al}_{\textnormal{3}}\textnormal{Li}$}

\newcommand{\vtesla}{Nvidia Tesla Volta V100}

\newcommand{\transcoder}{paraprobe-transcoder}
\newcommand{\synthetic}{paraprobe-synthetic}
\newcommand{\ranger}{paraprobe-ranger}
\newcommand{\surfacer}{paraprobe-surfacer}
\newcommand{\spatstat}{paraprobe-spatstat}

\newcommand{\araullo}{paraprobe-araullo}
\newcommand{\fourier}{paraprobe-fourier}
\newcommand{\indexer}{paraprobe-indexer}
\newcommand{\intersector}{paraprobe-intersector}

\begin{document}
\title{On Open and Strong-Scaling Tools for Atom Probe Crystallography: High-Throughput Methods for Indexing Crystal Structure and Orientation}
\author[1]{Markus K\"uhbach}
\author[2]{Matthew Kasemer}
\author[1,3]{Baptiste Gault}
\author[1,4]{Andrew Breen}
\affil[1]{Max-Planck-Institut f\"ur Eisenforschung GmbH, Max-Planck-Stra{\ss}e 1, D-40237 D\"usseldorf, Germany}
\affil[2]{Department of Mechanical Engineering, The University of Alabama, Tuscaloosa, AL 35487, United States}
\affil[3]{Department of Materials, Imperial College London, Royal School of Mines, London, United Kingdom}
\affil[4]{now at The University of Sydney, Australian Centre for Microscopy \& Microanalysis, NSW 2006 Sydney, Australia}
\date{}
\setcounter{Maxaffil}{0}
\renewcommand\Affilfont{\itshape\small}

\maketitle

\begin{abstract}
Volumetric crystal structure indexing and orientation mapping are key data processing steps for virtually any quantitative study of spatial correlations between the local chemistry and the microstructure of a material. For electron and X-ray diffraction methods it is possible to develop indexing tools which compare measured and analytically computed patterns to decode the structure and relative orientation within local regions of interest. Consequently, a number of numerically efficient and automated software tools exist to solve the above characterisation tasks. 

For atom probe tomography (APT) experiments, however, the strategy of making comparisons between measured and analytically computed patterns is less robust because many APT datasets may contain substantial noise. Given that general enough predictive models for such noise remain elusive, crystallography tools for APT face several limitations: Their robustness to noise, and therefore, their capability to identify and distinguish different crystal structures and orientation is limited. In addition, the tools are sequential and demand substantial manual interaction. In combination, this makes robust uncertainty quantifying with automated high-throughput studies of the latent crystallographic information a difficult task with APT data.

To improve the situation, we review the existent methods and discuss how they link to those in the diffraction communities. With this we modify some of the APT methods to yield more robust descriptors of the atomic arrangement. We report how this enables the development of an open-source software tool for strong-scaling and automated identifying of crystal structure and mapping crystal orientation in nanocrystalline APT datasets with multiple phases.\footnote{This is a preprint of the paper submitted to \textit{Journal of Applied Crystallography.}}
\end{abstract}

\newpage
\section{Introduction}
Atom probe tomography (APT) is a powerful nanoscale analytical tool capable of reconstructing the 3D position and chemical identity of millions of individual atoms from a specimen \cite{Mueller1968,Blavette1993,Miller2000,Gault2012b,Larson2013a,Larson2013b,Lefebvre2016} with sub-nanometer resolution \cite{Kelly2007d,Kelly2009,Gault2009c,Gault2010c,Degeuser2020}. This unique capability makes APT a useful technique to study the 3D atomic architecture of solids and has provided invaluable scientific insight into fields such as materials science \cite{Hono1999,Herbig2014,Kuzmina2015,Chen2017}, geology \cite{Valley2014,Piazolo2016,Saxey2018}, semiconductors \cite{Castell2003,Giddings2018,Rigutti2018}, and even biology \cite{Gordon2011}.

The technique works by inserting a sharp, needle-shaped specimen with an end-tip radius of less than \SI{100}{\nano\meter} into an ultra-high-vacuum chamber ($<$\SI{1.4e-13}{\bar}). A standing voltage of a few \SI{}{\kilo\volt} is applied, on top of which either laser or high-voltage pulses are superimposed to induce time-controlled field evaporation of individual atoms from the surface. During these experiments the specimen is held at cryogenic temperatures in the range of \SIrange{25}{80}{\kelvin} to limit the influence of surface diffusion on the analysis. 

Accelerated by the electric field surrounding the specimen, the ions are collected by a position-sensitive and time-resolved detector. Information on the time-of-flight between the specimen and the detector enables the determination of the mass-to-charge ratio for each ion. These ratios are associated to the most likely elemental identity, i.e. atom type of each ion \cite{Hudson2011,Haley2017}. A reverse-projection algorithm, which uses the sequence of evaporation events, combined with the x and y detector hit positions of the ions, is used to reconstruct the 3D position of the atoms from the analysed specimen \cite{Bas1995,Geiser2009a,Gault2010d,Gault2011b,Hatzeglou2019,Fletcher2020}. The result is a 3D dataset of atom positions and atom type. Ultimately, crystallographic information may be determined from such data - known by the term \textit{atom probe crystallography} in recent literature \cite{Moody2011a,Araullopeters2012,Gault2012}.

APT datasets of crystalline materials often contain crystallographic information that is particularly useful for the calibrating of the tomographic reconstruction \cite{Gault2008,Gault2009a}, measuring crystallographic character of interfaces \cite{Breen2017}, and studying ordering \cite{Marquis2007b,Gault2012e,Bagot2017}. However, the crystallographic information is partially lost because of the limited spatial resolution, finite detection efficiency, and necessary simplifying assumptions made when applying back-projection algorithms \cite{Vurpillot2000a,Kelly2007d,Gault2009c,Gault2010c,Vurpillot2015}. In effect, these limitations result in noise which makes a retrieval and exploitation of crystallographic information with APT datasets difficult.

The experimental conditions and the physical properties of the material being analysed, affect the available crystallographic detail. Typically, pure metals such as aluminium and tungsten collected at low temperature contain the clearest crystallographic information \cite{Gault2010c,Gault2010b}. For more complex materials systems, this information becomes more difficult to observe; in particular when different phases give rise to locally varying evaporation field conditions \cite{Vurpillot2000a}. Increasing the base temperature of the analysis or using laser-pulsing mode to improve specimen yield has a detrimental effect on the quality of the crystallographic information that can be retrieved \cite{Gault2010c,Gault2010b}. This substantiates the need for robust methods and efficient implementations to give practitioners a tool for quantifying where a dataset contains how accurate crystallographic information.

A working strategy for extracting crystallographic information from datasets of single- and polycrystalline specimens is to evaluate the pattern in the hit densities in detector space. Such patterns are characterised by averaging a number of subsequent x and y detector hit positions from a few hundred thousand to a few million ions into a hit density pattern \cite{Moody2011a,Yao2016,Wei2018,Wei2019,Kuehbach2019a}. Such patterns form as a result of trajectory aberrations inherently related to the crystallography of the specimen and quantum effects \cite{Oberdorfer2013,Ashton2020}. The necessity to collect these patterns via integrating the signal over a substantial number of ions (through averaging over the entire effective detector area as well as in depth) makes this strategy spatially inaccurate. In addition, the detector space does not account for eventual spatial distortions along the main axis of the dataset. In effect, it is practically difficult to index different thermodynamic phases via differences of their crystal structures or back out more information on the shape of these phases or the (relative) orientation of their lattice.

On the contrary, this is the strength of electron diffraction methods. In particular, correlative electron diffraction methods performed on the specimen prior to running the atom probe experiment, such as Transmission Kikuchi Diffraction (TKD) \cite{Babinsky2014,Zaefferer2011,Keller2011,Trimby2012,Trimby2014,Sneddon2016,Breen2017,Schwarz2017} and transmission electron microscopy (TEM) techniques \cite{Herbig2018}, such as \textit{e.g.} nanobeam diffraction \cite{Herbig2014,Zhou2016} or high-resolution TEM \cite{Makineni2018,Liebscher2018}, are particularly useful in providing some crystallographic information, including crystal defects within the grains and at the grain or phase boundaries. Applying these correlative techniques, however, adds to further experimental complexity and presents challenges with respect to the alignment between the electron microscopy and the APT data \cite{Mouton2019}. 

Another strategy for reconstructing grain and phase boundaries implicitly within APT datasets is via analysing regions of preferred elemental segregation. In fact, when different atom types can serve as the markers, gradients in the nanochemistry can be detected via segmentation methods like isosurfaces \cite{Hellman1999}, an extraction of cell facets from tessellations \cite{Felfer2012b,Felfer2013}, or via artificial intelligence methods \cite{Zhou2020}. Without analysing the specific atomic arrangement at the interface, however, deducing the orientations of the adjoining crystals is an ill-posed task.

All the above arguments affirm the advantages of direct crystallographic measurements on the reconstructed atom positions within APT datasets. In many cases, latent crystallographic information is contained in such datasets but it is incomplete. At present, the lack of simple and efficient crystallographic processing tools means that such data goes largely underused. Given that crystallographic information is often incomplete, it is essential to quantify at which locations in the dataset it is available, how accurate and precise it is, and where such information is virtually not recoverable. Assessments of the same dataset with multiple of the above detailed crystallographic methods, i.e. heads-up, can help clarifying when analysing the reconstructed atom positions is substantiated, or when correlative microscopy, elemental segregation, or patterns in the detector space are the last resort.

Above analysis tasks do not considerably differ conceptually or fundamentally to the reconstruction of grains from atomic positions monitored in molecular dynamic (MD) simulations \cite{Stukowski2012,Larsen2016,Hoffrogge2017}. However, APT datasets display, in most cases, a stronger positional noise as is evidenced in MD simulations or expected from thermal lattice vibrations alone \cite{Lonsdale1948,Vurpillot2000a}. In addition to thermal lattice vibrations and eventual diffusion over the specimen surface prior launch, reconstructed APT datasets inherit inaccuracies such as missing atoms (\SIrange{10}{30}{\percent}) because of limited detector efficiency.

This explains the motivation in the past for developing specific approaches for APT data to mitigate the above challenges \cite{Vurpillot2003,Moody2011a,Moody2014,Breen2015}. Despite their success for single crystals, though, the methods and their implementation in software tools present practical limitations that justify further research on highly localised 3D orientation mapping methods. Above methods can also be tedious to employ and potentially inefficient for larger datasets and mapping approaches \cite{Wallace2018}. Not only do many tools demand substantial manual interaction, but also have no sophisticated software parallelisation, resulting in unnecessarily long analyses.

In addition, we observe that the landscape with respect to numerical methods and computational hardware has improved substantially over the past decade, bringing a new opportunity for computationally intensive structural and crystallographic analysis \cite{Favrenicolin2011,Katnagallu2017,Dibernardo2018,Kuehbach2020a} that was not practical in the earlier works on 3D atom probe crystallography \cite{Camus1995,Cerezo1998,Vurpillot2001}. Many- and multi-core central processing units (CPUs) and general purpose graphics card coprocessors (GPGPUs, accelerators, or GPUs for short) are now readily available. This has made orders of magnitude more processing power available to microscopists and microanalysts. These observations motivate this study.

Therein, we aim at closing several gaps in atom probe crystallography: First, we generalise existing atom probe crystallography methods to make these more robust tools for indexing arbitrary crystal structures. Next, we detail how these results can be used to develop an automated method for backing out respective orientations. We analyse specifically the robustness against positional noise and missing atoms. Next, synthetic datasets with a low and a high complexity of the grain and phase boundary network will be assessed to verify the methods. Thereafter, we assess application examples on experimental APT datasets to identify the capabilities and quantify the limitations of such methods. Finally, we connect how this work embeds into the larger picture of indexing crystal structure and mapping orientation via methods of the electron and X-ray diffraction community \cite{Campbell1998,Kolb2006,Maia2011,Lenthe2019,Hielscher2019}. We implement the numerical tools for above research as an open-source software with specific parallelised algorithms (for CPUs and GPUs). In effect, this delivers a tool for more rigorous, orders of magnitude faster, and better reproducible high-throughput studying for atom probe crystallography.

\section{Computational methods}
\subsection{General procedure for indexing crystal structure and backing out orientation}
We perform all analyses by scanning the dataset volume with a nanometer-sized (spherical) region of interest (ROI). This yields a collection of ROIs. Each local analysis for a ROI has two steps: In the first step, at least one, so-called (crystallographic) signature, is computed from the positions of selected atom types in the ROI. A signature encodes whether there is a long-range periodic arrangement of selected atom types along particular (crystallographic) directions. Eventually, multiple signatures (for different atom types) are computed per ROI to help distinguish different crystal structures. The signatures are images, whose formatting is a function of the signature detection methods. In the second step, we use the signature(s), for each ROI, to index the most likely matching crystal structure, or candidate for short, and back out the equivalent rotations that bring the encoded crystallographic directions in the signatures into a consistent alignment with the laboratory coordinate system. 

Signatures are computed with two established, but here modified, methods from atom probe crystallography - either the method from Araullo-Peters \textit{et al.} \cite{Araullopeters2015} or the method by Vurpillot \textit{et al.} \cite{Vurpillot2001}. The discussion in this paper is focused on the first method. Novel is that we modified this method to achieve not only more robust signatures than those computed in the original paper but also that we detail, for the first time for APT, how to back out systematically orientations with these signatures. Based on the reference space in which the two methods operate, we refer to the first method as the real space method (RSP) and the second as the reciprocal space method (FSP). 

\subsection{Methods for detecting signatures of long-range periodic atomic arrangement}
\begin{figure}[!htb]
\centering
\includegraphics[width=1.0\textwidth]{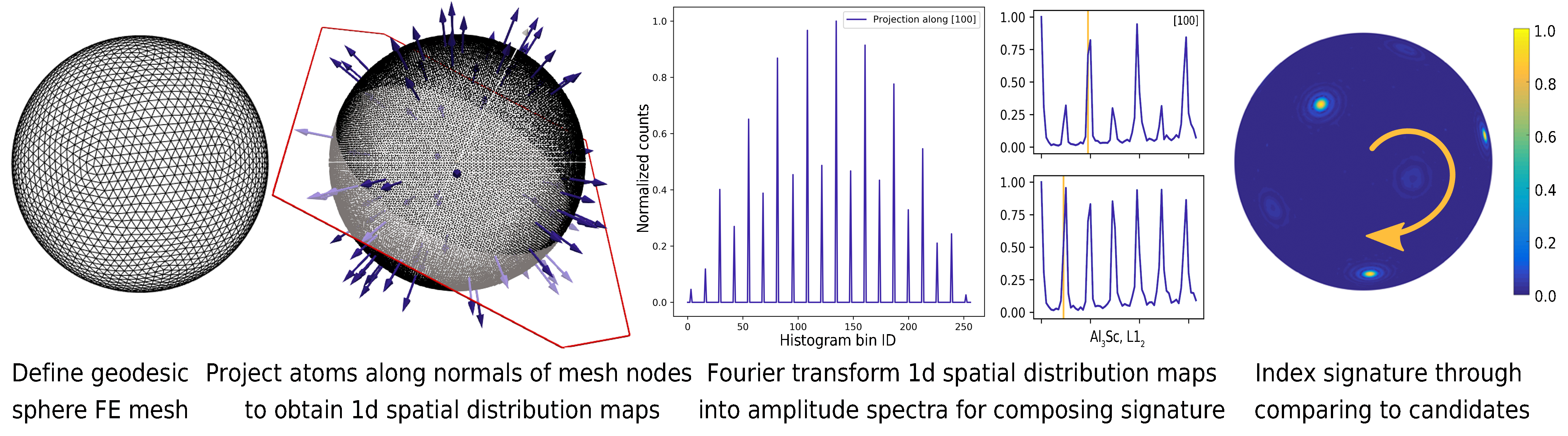}
\caption{In the real space method we compute signatures for each ROI and reference signatures of crystal structure candidates to compare these for indexing. The computation has several steps: First, the definition of one geodesic mesh which is used for all ROIs and defines the corresponding projection directions. Second, the computation of the one-dimensional spatial distribution maps (SDMs) for each direction and ROI. Third, the fast Fourier transformation of each SDM plus subsequent signal extraction using the modified strategy for extracting crystal-structure-specific peaks from the amplitude spectra to compose at least one signature per ROI. The fourth step is the indexing of the signatures by comparing each signature against a set of rotated reference (signatures for crystal structure candidates). The colour bar to the right shows normalised intensities.}
\label{fig:acqidx_workflow}
\end{figure}

\paragraph{Real space method (RSP)}
Beginning from the work of Araullo-Peters \textit{et al.} \cite{Araullopeters2015}, we project inter-atom-distances along a set of directions defined {\textit{a priori}} \cite{Araullopeters2015,Haley2019b}. The key steps are shown in Fig. \ref{fig:acqidx_workflow}. 

Each projection yields a one-dimensional spatial distribution map (SDM) \cite{Geiser2007,Moody2009a,Moody2009b}, and hence a set of histograms of projected distances for each ROI. A right-handed Cartesian coordinate system (laboratory) is assumed whose xy-plane is located at the base of the dataset. 
Each local analysis yields a set of (projection) directions, encoded as elevation-azimuth pairs. We propose to align these directions with the outer unit normals of the nodes of a geodesic sphere finite element mesh \cite{Popko2012,Pokhrel2018}. In this work, the finite element (FE) mesh contained $N_v = \SI{40962}{}$ vertices. For each direction, i.e. one-dimensional SDM, we compute one histogram using Eq. \ref{eq:projection}. 

\begin{equation}
b_i(\alpha, \beta) = \lfloor{\frac{({\begin{bmatrix} cos(\alpha)cos(\beta) \\ cos(\alpha)sin(\beta) \\ sin(\alpha)\end{bmatrix} \cdot (\bm{p_i} - \bm{p_c}) + R + \Delta R) \cdot 2^m}}{\Delta R+2R+\Delta R}}\rfloor
\label{eq:projection}
\end{equation}


Therein, $R$ is the radius of each ROI, while $\alpha$ and $\beta$ denote the elevation and azimuth, respectively. Histograms have a total of $2^m$ bins, with $m$ an integer in the range \SIrange{8}{12}{}, and $\Delta R$ the bin width. Lower case bold letters specify vectors in 3D. For each $c$-th ROI, the position $\bm{p}_i$ of the $i$-th atom is evaluated relative to the centre of the ROI $\bm{p}_c$. Projected distances are cast into a bin ID $b_i$. 

Next, we compute a fast Fourier transformation of each histogram and inspect the resulting amplitude spectrum for each direction. Provided that the lattice planes are aligned nearly perpendicular to a particular (projection) direction $\bm{v}$, we expect to find peaks in the amplitude spectrum. These peaks should correlate with the spacing of particular distances between lattice planes or multiples of these distances. Consequently, a spherical image can be composed in such a way that it is a signature of the corresponding lattice periodicity signal for each direction within the ROI. By virtue of construction, each image also encodes the relative orientation of the crystal volume.

The key modification to the original work \cite{Araullopeters2015} is in the choosing of particular bins instead of arbitrary bins of the amplitude spectrum to compose the signature. Repeating this choice for all directions results in said spherical image ${\mathcal{S}}^{msr}_c$ for each ROI $c$. When computed from a dataset that we wish to index, we refer to the signatures also as measured signatures in order to distinguish them from the reference signatures that we explain in the next sub-sections.

The key advantage of this approach is that this computation is independently executable for each direction and each ROI. This independence brings substantial potential for data parallelised execution. To the best of our knowledge no implementation has exploited this advantage so far. A detailed analysis of the numerical costs of the RSP method is reported in the supplementary material.

\paragraph{Reciprocal space method (FSP)}
Following Vurpillot \cite{Vurpillot2001}, we implemented a second method for computing signatures. That is to evaluate Eq. \ref{eq:directft} to compute a direct Fourier transform of the positions for all atoms $N_w$ of a selected atom type within each ROI $c$. Here, $w$ is a counting variable, $i$ is the imaginary unit, and $\bm{k}$ a reciprocal space position vector.

\begin{equation}
\hat{F}(\bm{k}) = \sum^{N_w} exp(-2i\pi \bm{k} \cdot ({\bm{p}}_w - {\bm{p}}_c))
\label{eq:directft}
\end{equation}

Contrary to signatures for a ROI obtained with RSP, i.e. spherical images, the reciprocal space method yields the signatures as three-dimensional image stacks. The reciprocal space method also allows for substantial data parallelised execution, which we detail in the supplementary material. However, a key difference to the RSP method is that this requires more arithmetic operations per ROI because there are typically far more reciprocal space grid points ($\bm{k}$) to compute than directions ($N_v$). So far, this has substantially restricted the application of the reciprocal space method until recently \cite{Katnagallu2017,Dibernardo2018} when the specific performance of GPUs became higher and cheaper. 

\subsection{Extraction of the crystal structure and orientation}
Both RSP and FSP yield signatures of crystallographic information for each ROI. At least two strategies exist to identify now the crystal structure and orientation from these signatures: Either we restrict ourselves to a particular small set of candidate crystal structures (candidates, for short) or we do not make such simplification and try to test against all possible crystal structures. Here, we detail a solution in accord with the first strategy. 

Thus, it suffices to compare, for each ROI, each signature against a set of reference signatures for all candidates and rotated versions of the reference signatures. The strategy is similar to the indexing of electron back-scatter diffraction data (EBSD) \cite{Schwartz2009} whereby a measured Kikuchi pattern is indexed with a predicted Kikuchi pattern while assuming a few different crystal structures as candidates. The best match is then quantified via suitable descriptors \cite{Wright2015}. 

Likewise, here we compare, for each ROI, the collected signatures against reference signatures and rotated versions of these for a set of crystal structure candidates. The reference signatures for the candidates, or reference(s) for short, are characterised for synthetic single crystals with a defined atomic arrangement and defined noise. Specifically, the references are computed as atom-type-specific spherical images (for RSP) from synthetic single crystal datasets. The resulting spherical images are rotated to sample the orientation space \cite{Bunge1982,Rowenhorst2015,Bachmann2010b}. Specifically, the orientation set $\mathcal{G}$ contained approximately $\SI{0.62}{}$ million orientations (also referred to as test orientations) with $\frac{4}{m}\bar{3}\frac{2}{m}$ crystal symmetry and \SI{1}{\degree} angular spacing. Further details are reported in the supplementary material.

\subsection{Indexing}
With the above assumptions and definitions, the task of indexing the crystal structure and orientation decreases for each ROI and crystal structure candidate to a comparison of at least one spherical image, i.e. signature to a set of rotated spherical images, i.e. rotated reference (signatures). To accomplish this, one can either evaluate the image intensities as a whole, via e.g. cross-correlation, or register the images by matching against spots of high image intensity. 

The indexing algorithm works as follows: First, we normalise the image intensities of the signatures for the ROIs and for the references for each candidate. Second, we identify the locations and intensities of a predefined number of the highest absolute image intensities for each candidate. Third, we build a lookup table which guides where to probe nodal values of the signature to compare implicitly against all rotated references for the orientation set $\mathcal{G}$. With this, we evaluate the signatures for each candidate and ROI at the precomputed image positions to quantify how closely signatures and references match. Further numerical details and costs of this approach are described in the supplementary material.

\subsection{Implementing these numerical methods into a software toolbox}
\paragraph{Defining a workflow for indexing crystal structure and orientation}
We implemented the above methods as additional tools (\araullo{} for the RSP and \fourier{} for the FSP method, and \indexer{}) into \paraprobe{}. This software is an open-source toolbox for high-throughput analysing of APT datasets \cite{Kuehbach2019c,Kuehbach2020a}. The ROIs were either placed on the positions of a 3D grid or placed via random sampling. Specific methods \cite{Kuehbach2020a} assured that only ROIs within the dataset were analysed. Furthermore, we implemented a proof-of-concept to an iterative grid refinement to allow detailed analyses at a much finer spatial resolution of the ROI grid without having to waste computational resources on locations where the signal quality is lower. The analysis workflow is defined through a \python{} script, whose details are described in the supplementary material. Hands-on examples in the form of \jupyter{} notebooks are provided to guide experimentalists to apply the methods to their own datasets \cite{Kuehbach2020d}.

\paragraph{Parallel implementation of the software}
With several multi-core CPUs and GPUs typically on board, modern computers offer multiple layers of parallel resources \cite{Hennessy2012,Rauber2013}. To use all these resources productively, we built on previous work \cite{Kuehbach2020a} and parallelised for CPUs (real space method) and also GPUs (reciprocal space method). Using the Message Passing Interface (MPI) library, ROIs were split into sub-sets. These sub-sets were delegated round-robin to the computing nodes and distributed further on these via multithreading - realised with Open Multi-Processing (OpenMP) \cite{Chapman2007,Kuehbach2020a}. Atom positions were always queried using OpenMP. GPU instructions were implemented through Open Accelerator (OpenACC) compiler directives and CUDA library commands \cite{Nvidia2019}, respectively. Each GPU was instructed by an own MPI process with an own OpenMP master thread. 

The Hierarchical Data Format (HDF5) library was used \cite{HDF52020,Prabhat2014} to store all data and metadata transparently and performantly. Herewith, we signify our desire to remove unnecessary barriers with respect to the FAIR data stewardship principles \cite{Wilkinson2016,Draxl2020}. Specific sequential implementation tricks are detailed in the supplementary material.

All analyses were executed on the TALOS computer cluster \cite{Kuehbach2019c,Kuehbach2020a,Kuehbach2020e}. Each node has two Intel Xeon Gold 6138 twenty-core processors with access to \SI{188}{\giga\byte} main memory in total. Each node is equipped with two \vtesla{} \cite{Nvidia2017} GPUs with \SI{32}{\giga\byte} memory each. We used at most \SI{80}{} of the TALOS computing nodes and their GPU pairs. All resources were used exclusively and the elapsed time for accessing files and processing data was accounted for individually. Further details are reported in the supplementary material.

\section{Results}
\subsection{Verification of the methods}
\paragraph{Does the real space method yield signatures specific for a crystal structure?}
First, we verify that RSP yields peaks in the amplitude spectra at positions that are specific for the crystal structure and the orientation of the crystal. This is a requirement for a reliable distinguishing between different crystal structures. Synthetic datasets were created for this purpose as these assure a rigorously controlled atomic architecture. Specifically, we synthesised three cylindrical datasets with a height-to-radius of $H = 2R$ and a total of $\SI{25e6}{}$ atoms:

\begin{itemize}
    \item A face-centred cubic (f.c.c.) aluminium lattice with $a_{Al} = \SI{4.05}{\angstrom}$, COD ID 9008460, according to the Crystallography Open Database (COD) \cite{Downs2003,Grazulis2009}
    \item A $\textnormal{L1}_{\textnormal{2}}$ structure \althreesc{} phase lattice with $a_{Sc} = \SI{4.10}{\angstrom}$ according to \cite{Villars2016a}.
    \item A body-centred cubic (b.c.c.) tungsten lattice with $a_{W} = \SI{3.16}{\angstrom}$, COD ID 9008558.
\end{itemize}

All datasets represent single crystals (defect-free and of single-phase). Two instances were created for each of the three datasets. For the first instance the lattice remained unrotated, thereby representing a single crystal in \bunge{0.0}{0.0}{0.0} orientation using the Bunge-Euler notation. For the second instance, the single crystal was rotated \bunge{8.0}{8.0}{8.0}. First we work with the single crystals with unrotated lattice. We placed a single ROI ($R = \SI{20}{\angstrom}$, $m = 8$) in the centre of the dataset and projected the atoms in the ROI along the $26$ symmetric variants of the \hkl<100>, \hkl<110>, and \hkl<111> directions.

\begin{figure}[!htb]
\centering
\includegraphics[width=0.9\textwidth]{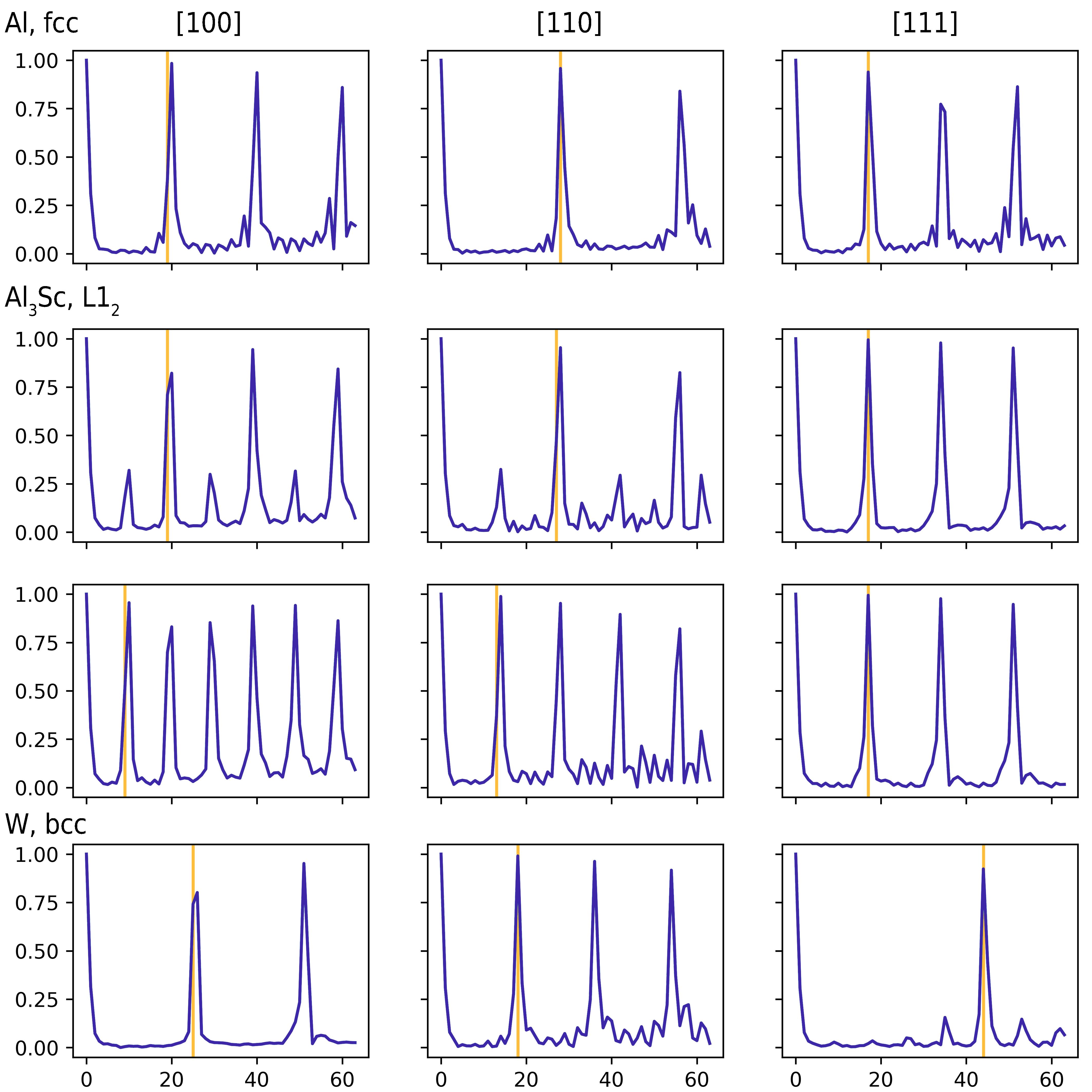}
\caption{Verification that signatures can distinguish different crystal structures and encode different orientations. Compared are the low frequency part of the one-sided FFT amplitude spectra for the aluminium, the \protect\althreesc{}, and the tungsten synthetic datasets after projecting along specific crystallographic directions (\hkl[110], \hkl[110], and \hkl[111]). The two rows in the middle display the results for the ${\textnormal{L1}}_{\textnormal{2}}$ crystal structure (with the aluminium sub-lattice in the upper and the scandium sub-lattice in the lower middle row). The x axis shows amplitude spectra bin IDs. The vertical orange lines mark the theoretical lattice plane spacing for lattice planes which are stacked perpendicular to the respective crystallographic (projection) directions. Exemplified for the b.c.c. tungsten lattice, we expect to find an alternating sequence of \hkl<100> and \hkl<200> planes with $0.5a_W$ spacing and an equal planar density of the tungsten atoms for the two inspected crystallographic plane sets. Assume the signal length is $\mathcal{L} = 2^m$ with $m = 8$. The sampling frequency is $f_s = \frac{\mathcal{L}}{2(R+\Delta R)}$ with $\Delta R = \frac{R}{2^{m-1}-1}$. Now for $R = \SI{2.0}{\nano\meter}$ and a reciprocal spacing $f = \frac{1}{0.5a_W}$, we can verify that the amplitude peaks in the $19$-th bin ($b = \lfloor{\frac{f\cdot\mathcal{L}}{f_s}}\rfloor$).}
\label{fig:verify_radius}
\end{figure}

Figure \ref{fig:verify_radius} summarises the results by comparing selected one-sided amplitude spectra of the fast-Fourier-transformed SDMs. 

Specifically, we compare amplitude spectra for all three crystal structures (rows of the image matrix) and selected specific crystallographic directions (columns of the image matrix). The results are representative for other analysed ROI radii and frequency resolutions. The figure documents that the modified signal selection strategy yields amplitude spectra with distinct peaks. There is always a peak at the origin which accounts for the total number of atoms for the analysed type in the ROI. Further peaks in different bins are detected. Their bin position encodes the spacing of a stack of lattice planes. As confirmed by the vertical orange lines, explained in further detail in the figure caption, the positions of the peaks match theoretical expectations.

When comparing the peaks for aluminium versus tungsten in the amplitude spectra for the \hkl[100] direction for instance, it is reassuring to find that the peaks are in different bins because the lattice constant of aluminium is different to that of tungsten. The two rows in the middle of the figure depict how the peaks of the specific $\textnormal{L1}_{\textnormal{2}}$ structure candidate differs from the signature for the aluminium lattice, although the dimensions of both unit cells are almost the same.

We learn that to distinguish between these crystal structures, if they were to exist in the same dataset, it is necessary to evaluate two amplitude spectra: one for the atoms of the aluminium sub-lattice and another one for the atoms of the scandium sub-lattice. We also learn that observing a single peak in an amplitude spectrum characterises the spacing between a specifically oriented single stack of lattice planes. In the examples above e.g. lattice planes with normals \hkl[\pm u00], with $u$ as an integer. 

Consequently, these peaks can be used to compose three-dimensional signatures of the respective crystal structures which are specific for a given set of lattice planes including all symmetric variants \hkl{hkl}. For cubic crystal symmetry this implies that the signatures are capable to detect equivalent \hkl<uvw> directions. It is possible to compose a signature from peaks at different locations and use a colouring scheme to distinguish the peaks. 

In effect, the real space method yields distinct signatures for a crystal structure. The \althreesc{} case shows that if two structures have strong similarities, it is necessary to study a combination of signatures that is computed from different atom types and evaluating their corresponding sub-lattices.

Our procedure to specify the relevant bins in the amplitude spectra is more robust than the strategy of the original authors --- see section 2.2, bullet point 4c of the original paper \cite{Araullopeters2015}. Therein, the authors propose to pick a peak at the bin which is closest to where the amplitude spectrum has its median amplitude. However, as we discussed, each bin represents a specific spacing. In effect, such a peak selection rule cannot guarantee that for each projected direction and ROI always the same bins get analysed.

\paragraph{How robust are signatures from RSP against noise?}
Before the computed signatures can be used for indexing, we need to address their robustness against noise because the reconstruction of an APT dataset from detector hit and time-of-flight measurements faces the challenges of undetected ions, trajectory aberrations, and trajectory overlap \cite{Larson2013a,Vurpillot2015,Devaraj2018}. Therefore, the effect of noise from finite detection efficiency, i.e. missing atoms in the reconstructed dataset, and positional noise, i.e. imprecisely placed atoms, needs to be accounted for. The spatial resolution within an atom probe reconstruction is anisotropic and typically higher along the local normal to the specimen surface than in the local tangent plane. For this purpose, we proceed by building copies of the rotated aluminium synthetic datasets from the above verification. These datasets were modified to create two types of noisy copies:
\begin{enumerate}
    \item Atoms were kept at their position but partially removed in a spatially random manner so that $N_{f} = \eta N_{i}$ atoms remain, with $\eta = 1.0$, $0.75$, $ 0.50$, and $0.25$.
    \item Atoms were not removed but displaced by applying an anisotropic Gaussian displacement kernel. The standard deviation of the kernel was always $\sigma_x = \sigma_y = \sigma_{xy} = 2\sigma_z$ but different $\sigma_z =$ $0.00$, $0.25$, $0.50$, $0.75$, $1.00$ and $\SI{2.00}{\angstrom}$ were probed, respectively.
\end{enumerate}

In addition to contributions from noise, rigorous analyses also need to take into account that placing the ROIs in the dataset is random sampling. In effect, we expect that the nodal intensities of the signatures scatter statistically, especially when using ROIs with a few angstrom radius only. To quantify such scatter, we work with a statistical ensemble of \SI{1e4}{} ROIs for each synthetic dataset. The results are documented in Fig. \ref{fig:verify_noise_positional} and in the supplementary material.

\begin{figure}[!htb]
\centering
\includegraphics[width=0.5\textwidth]{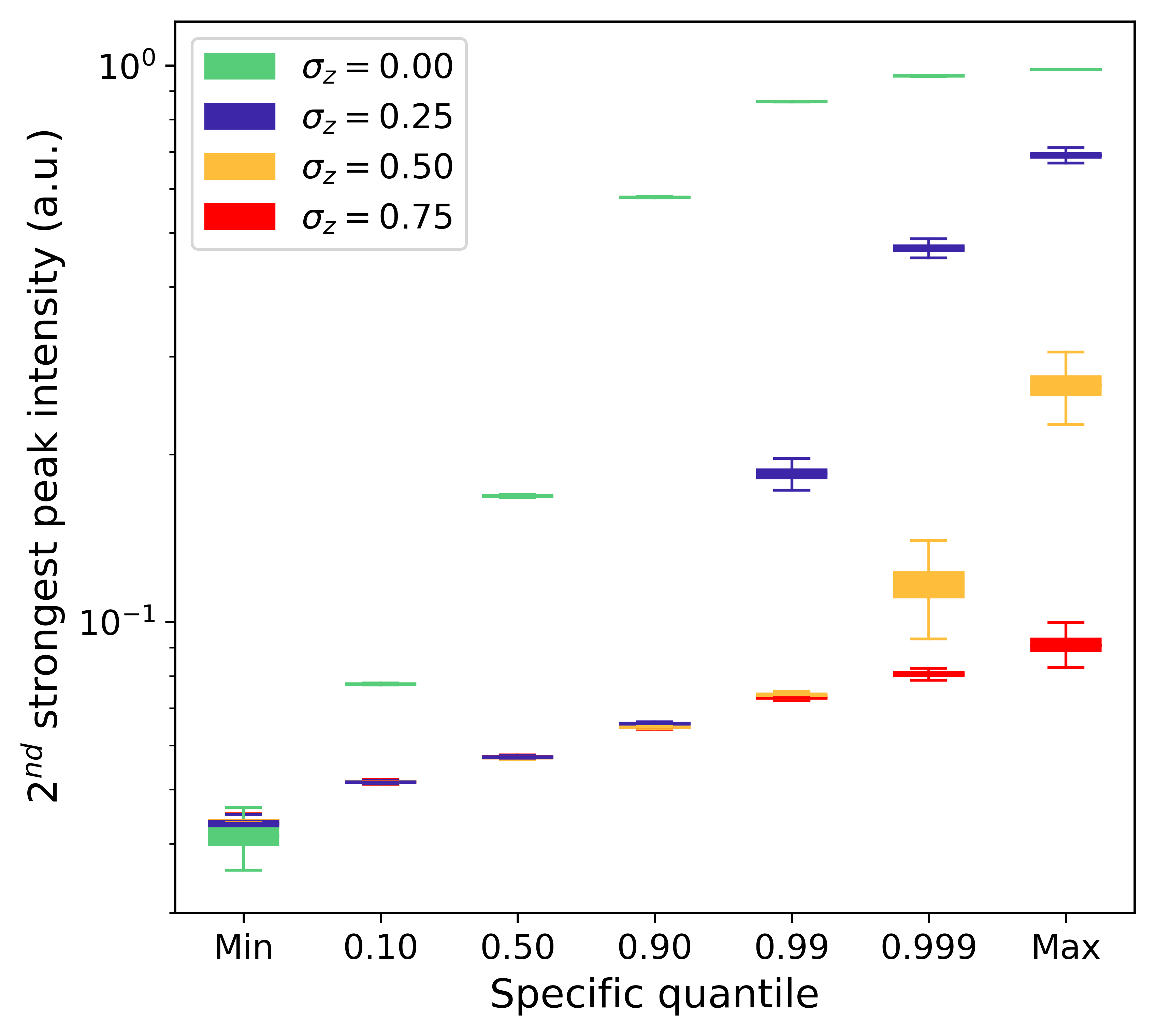}
\caption{Quantification of the real space method against positional noise. For each ROI, we identified the peaks in all amplitude spectra and report the individually second strongest peak per amplitude spectrum. With $N_v$ FE mesh nodes, i.e. $N_v$ directions (or corresponding SDMs), this yields one cumulative distribution per ROI. Next, specific quantiles of the distribution were extracted for each ROI and displayed for the entire ROI ensemble. This condenses how the results differ for all amplitude spectra ($40962$ per ROI) and all ROIs ($10000$ in total). We repeat this statistical analysis for all signatures from the datasets with different atom displacement strengths ($\sigma_z$ ) and compare. Contrary to noise from missing atoms, a substantial reduction of the signal-to-noise ratio is observed the stronger the atoms are displaced.}
\label{fig:verify_noise_positional}
\end{figure}

The figures display descriptive statistics for selected quantiles of the distribution of the second strongest peaks. Each amplitude spectrum contributes one of the $N_v$ peaks for a ROI. Each ROI contributes one quantile value. We report the intensities of the second strongest peak because the strongest peak in the (unnormalised) amplitude spectrum gives only the total number of atoms inside the ROI. We document in the supplementary material that the real space method yields signatures whose signal-to-noise ratio does not reduce substantially when removing atoms randomly.

For an increasingly stronger displacing of the atoms, though, Fig. \ref{fig:verify_noise_positional} documents a substantial reduction of the signal-to-noise ratio. For a standard deviation of $\sigma_z = \SI{0.25}{\angstrom}$ along the dataset main axis, the resulting displacements already exceed those of thermal lattice vibrations \cite{Lonsdale1948}. Nevertheless, the peaks remain strong against the background. Even for $\sigma_z = \SI{0.5}{\angstrom}$ the intensity peaks remain detectable but at half the signal-to-noise. In this case already half of the atoms are displaced by more than \SI{20}{\percent} of the lattice constant. For even stronger displacements the peaks eventually disappear, when approximately \SI{20}{\percent} of the atoms are displaced statistically by distances of more than half the unit cell. These results suggest that the signatures show still distinct intensity peaks even for datasets with displacements which are stronger than thermal lattice vibrations. This offers potential for automatic indexing and orientation mapping. 

We have implemented such analyses for arbitrary space groups. We can learn from the examples above that the key quantities to inspect when distinguishing different crystal structures in a noisy dataset is how the position deviations (due to noise) compare to the distribution of the nearest and the higher-order nearest neighbour spacing of the respective atom types in the unit cell. 

\subsection{Verifying the indexing for synthetic data and single crystals}
Consequently, we take the next step of the verification and attempt the indexing of a single crystal. Again synthetic datasets are used because these can be created in arbitrary orientation, and thus enable a quantitative assessment of how precisely and accurately orientations are identifiable. Specifically, four single-crystalline aluminium datasets were created. All datasets represent a lattice with approximately $\SI{20e6}{}$ atoms in orientation \bunge{8.0}{8.0}{8.0}. Different displacements of atoms ($\sigma_z = \SI{0.00}{}$, $\SI{0.25}{}$, $\SI{0.50}{}$, and $\SI{0.75}{\angstrom}$, respectively) were probed. A total of $\SI{1e4}{}$ ROIs were placed randomly.

\begin{figure}[!htb]
\centering
\includegraphics[width=0.5\textwidth]{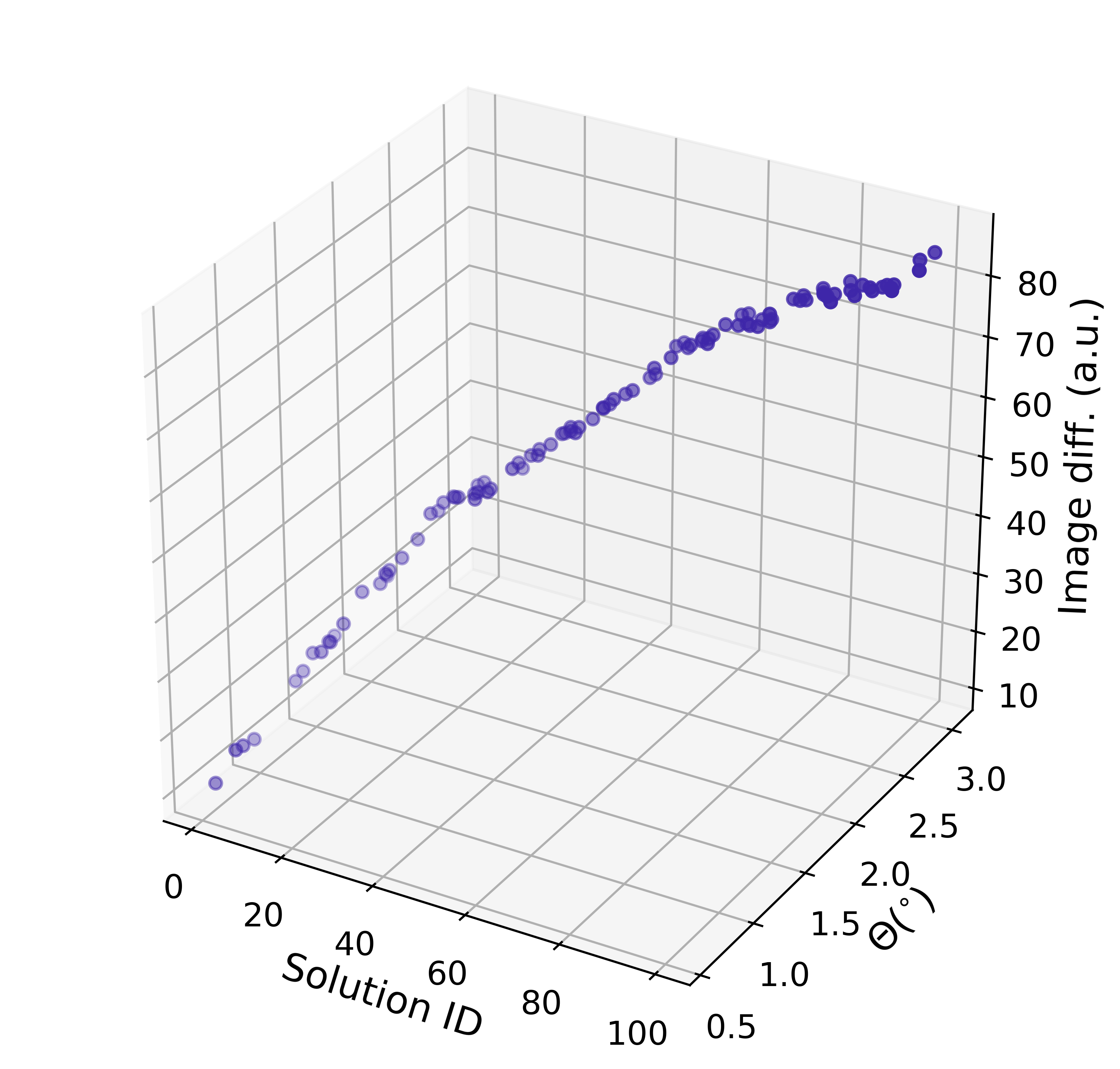}
\caption{Verification of the indexing for a single ROI and the synthetic single crystal. It is shown the image difference (image diff.) of the $100$ best matching solutions as a function of the solution ID and the disorientation angle. Best matching in this context means rotated versions of the reference that match with the signature of the ROI. The disorientation is computed between the orientation which the rotated version of the reference represents and the true orientation of the single crystal.}
\label{fig:verify_sxindexing_single}
\end{figure}

Figure \ref{fig:verify_sxindexing_single} documents the key results when indexing the signature of a single such ROI against a single candidate: an image difference value (image diff.) which quantifies how strongly the signature of the ROI differs from a particularly rotated reference (signature) of the candidate. Lower values indicate a better matching of the image intensity peaks. Given that in this verification we prescribe the orientation of the single crystal, it is possible to compute the difference between the orientation represented by the particularly rotated reference and the true orientation of the single crystal. 

The results document that \paraprobe{} backs out the true orientation precisely, accurately, and consistently. In fact, the solution with the lowest ID has the lowest image difference and the lowest disorientation angle ($\Theta$). Our method also recovers that multiple similar well-matching solutions exist. All of which register rotated references, representing the crystal structure candidate in different orientations. Some of these have a slightly higher disorientation angle (\SIrange{1}{3}{\degree}); suggesting that these are slightly rotated signatures of the true orientation or its symmetrically equivalent variants, respectively.

\begin{figure}[!htb]
\centering
\includegraphics[width=0.5\textwidth]{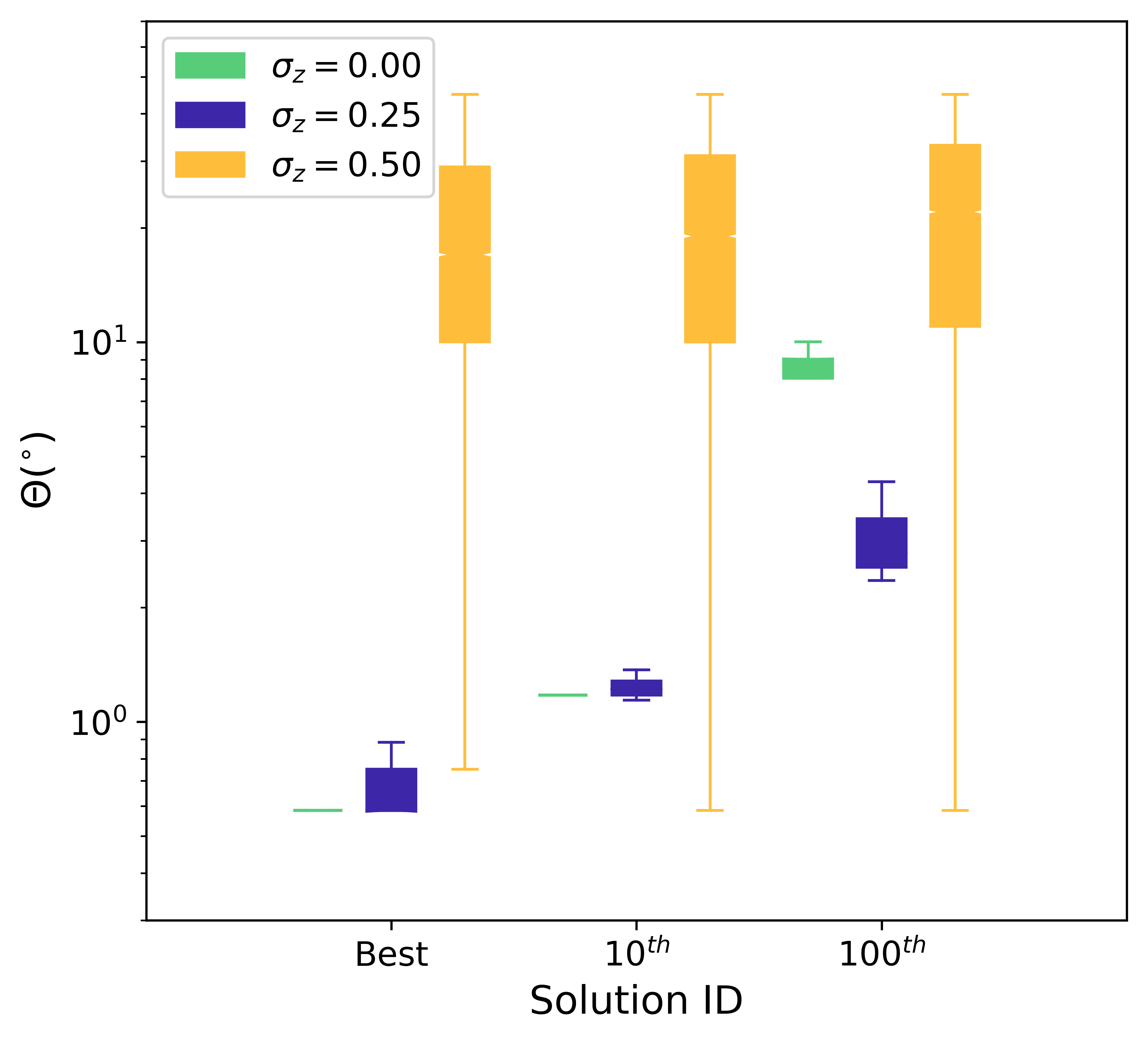}
\caption{Verification of the indexing for all the ROIs of the synthetic single crystal. The best and two less well-matching solutions are picked for each ROI and characterised with respect to their disorientation to the true orientation of the single crystal. Stronger disorientation indicates poorer indexing. Stronger scatter indicates less robust indexing. The results confirm that it is possible to index robustly for up to $\sigma_z = \SI{0.25}{\angstrom}$, $\sigma_x = \sigma_y = 2\sigma_z$ displacement (standard deviation).}
\label{fig:verify_sxindexing_stats}
\end{figure}

Having discussed a single ROI, we focus next on the ROI ensemble and investigate if the true orientation is recoverable in all regions of the single crystal. Figure  \ref{fig:verify_sxindexing_stats} confirms that \paraprobe{} recovers this information. 

The figure summarises statistics for all ROIs. The results answer how the particular well-matching solutions scatter for a given amount of positional noise. Again, the disorientation angle between the known solution and the calculated solution in the ROIs quantifies the indexing quality. The disorientation should ideally be the same for each ROI and close to the resolution of the orientation grid $\mathcal{G}$ ($\approx\SI{1}{\degree}$) because of probing a single crystal. The results signify that the method indexes correctly because strongly disoriented best solutions are not found. It is reassuring that the poorer solutions have a consistently higher disorientation angle. We also observe that indexing is possible as long as the noise remains below $\sigma_z = \SI{0.5}{\angstrom}$ (in this example). 

\subsection{Verifying the indexing of polycrystals}
As the last verification, we attempt indexing a synthetic polycrystal. For this purpose we built a needle-shaped synthetic dataset with approximately $\SI{200e6}{}$ atoms \cite{Kuehbach2019a,Kuehbach2020a}. Grains with an average spherical equivalent diameter of $\SI{200}{\angstrom}$ were created by placing seed points of a three-dimensional Poisson-Voronoi tessellation \cite{Okabe2000} inside the dataset. After assigning a random orientation for each grain, we filled each corresponding Voronoi cell with a local aluminium lattice and a (random) orientation for this lattice.

The practical advantage in this verification study is that the shape of the grains is rigorously defined. This enables us not only to compute the location of each boundary between any two cells (grains) but also to compute the location of the junctions between the interfaces. This offers the unique opportunity to quantify how much volume of each ROI lays within a particular grain (Voronoi cell). We define this volume fraction as $\zeta_k$, i.e. how large is the volume fraction which grain $k$ occupies of a ROI. Values of $\zeta_k = \SI{1.0}{}$ encode that the ROI is completely embedded in grain $k$. A value of $\zeta_k = \SI{0.5}{}$ means that only half of the ROI volume is covered by grain $k$. Three realisations of this dataset were created, differing only in their positional noise ($\sigma_{z} = \SI{0.00}{}$, $\SI{0.25}{}$, and $\SI{0.50}{\angstrom}$), respectively but using the same seeds for the grains. We scanned a 3D grid of ROIs with ${(\SI{10}{\angstrom})}^3$ spacing. 

\begin{figure}[!htb]
\centering
\includegraphics[width=1.0\textwidth]{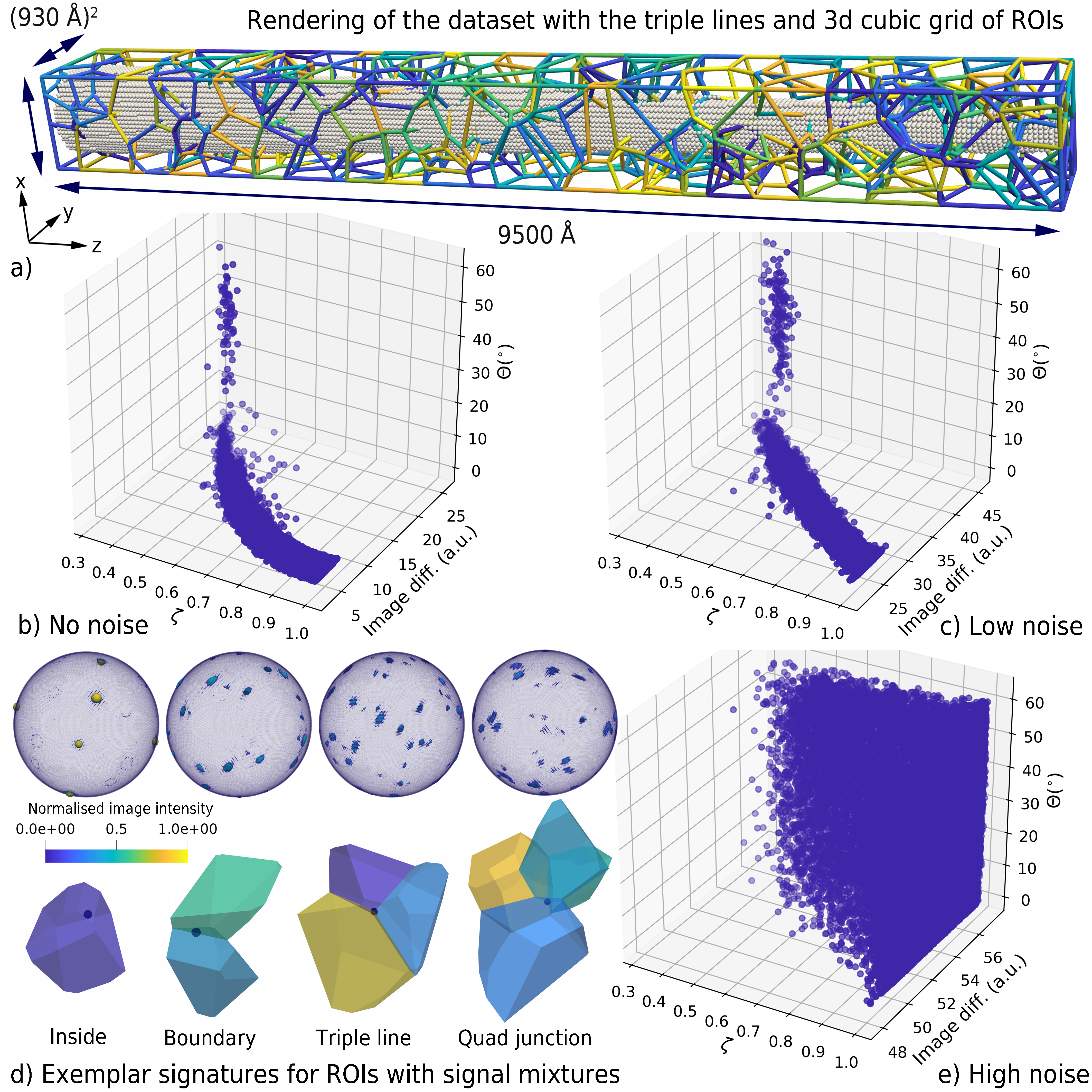}
\caption{Final verification with a synthetic polycrystal. The ROIs are shown as grey spheres. The wireframe in a) represents the triple lines between the grains (Voronoi cells) colored by grain ID. Figures b) ($\sigma_z = \SI{0.00}{\angstrom}$), c) ($\sigma_z = \SI{0.25}{\angstrom}$), and e) ($\sigma_z = \SI{0.50}{\angstrom}$) display for all ROIs of the dataset the disorientation to the individual true orientation of each ROI. The blue points display the disorientation (angle) as a function of the image difference and the signal contribution ($\zeta$) from the grain with the highest signal contribution. The results confirm that fully automated indexing is possible as long as the positional noise remains lower than \ensuremath{\sigma_z = \SI{0.5}{\angstrom}}. For ROIs at grain boundaries and junctions it is already evident from b) that the solution quality deteriorates systematically the stronger the neighbouring grains mix signal contributions into the signature.}
\label{fig:verify_px_indexing}
\end{figure}

Figure \ref{fig:verify_px_indexing} depicts the interface network of the polycrystal (coloured wire-frame) and the 3D ROI grid (grey spheres).
We use the notation and quantities that were introduced with Fig. \ref{fig:verify_sxindexing_single}. Figures \ref{fig:verify_px_indexing}b), c), and e) depict a collection of points. Each point represents the image difference (image diff.) for the closest matching orientation which \indexer{} suggests for each ROI. Each point shows one ROI. For each ROI we computed the strongest volume contributions $\zeta$ (on the x axis). The grain with the largest volume fraction defines the reference grain for the ROI. With this reference grain, it is possible to compute the disorientation angle between the orientation suggested by \indexer{} and the true orientation. Figures \ref{fig:verify_px_indexing}b), c), and e) show the solution quality (image diff., on the y axis) as a function of the disorientation (angle) to the true orientation (on the z axis) and the strongest volume contribution $\zeta$ (on the x axis). Now, one can compare the indexing success for no positional noise, Fig. \ref{fig:verify_px_indexing}b), to the results for a low,  Fig. \ref{fig:verify_px_indexing}c), and a high amount of positional noise, Fig. \ref{fig:verify_px_indexing}e), respectively. Figure \ref{fig:verify_px_indexing} summarises the key achievement of this work: it is possible to index APT datasets with fully automated methods, comparable to 3D orientation mapping for SEM/EBSD, provided that the reconstruction is sufficiently accurate. Most ROIs are solved accurately and precisely, many with better than $\SI{1}{\degree}$ angular resolution, thanks to the combination of the finite element mesh and orientation set ${\mathcal{G}}$.
 
This is an improvement compared to previous studies for several reasons: not only is it the first work to use a fully automated protocol for quantifying the effect of signal mixture rigorously. Our approach even works for specimens with $\SI{200e6}{}$ atoms and executes substantially faster because of sequential optimisation combined with parallelisation. This will be proven in the benchmark section. Neither do we rely on manual analyses \cite{Liddicoat2010} nor do we work in detector space \cite{Yao2016,Wei2018,Wei2019} exclusively or need to have an elemental segregation at the interfaces \cite{Felfer2015a,Zhou2020}.

Having verified the functioning and consistence of the tool, the results in Fig. \ref{fig:verify_px_indexing} document that the indexing of a dataset fails systematically beyond a certain amount of positional noise \cite{Vurpillot2001,Breen2015}. There are two contributions which act concomitantly to cause indexing failures:
\begin{itemize}
    \item The signal-to-noise-ratio decreases with increasing positional noise. Thereby, all possible peaks with which a signature is indexed get weaker. This reduces the discriminatory power of the image comparison method; and thereby the capability of the algorithm to identify as few as possible and still reliably which candidates match with the rotated references.
    \item The results document that indexing fails first close to interface junctions, i.e. for ROIs where the signal comes from multiple crystals. Selected examples for grain boundaries, triple lines, and higher-order junctions are shown in Fig. \ref{fig:verify_px_indexing}d).
\end{itemize}

An explanation for incorrect indexing is evident in Fig. \ref{fig:verify_px_indexing}b): We generated a nanocrystalline aggregate with a quasi random texture. Therefore, most grain boundaries are of high-angle character \cite{Mackenzie1958}. In effect, only one orientation of a grain pair decodes a low disorientation, while the other disorientation centres around the peak of the so-called Mackenzie distribution. The values for the standard deviations ($\sigma_{xy} = 2\sigma_z$ with $\sigma_z \leq \SI{0.75}{\angstrom}$) are comparable to the ones that were used in \cite{Vurpillot2001,Breen2015}.

\subsection{Assessing the significance of signal mixture}
These examples quantify how strongly a certain amount of signal mixture deteriorates the indexing quality. For reconstructed datasets from real APT experiments, making such rigorous comparison is very difficult without having access not only to correlative results in general but to reconstructions with angstrom precision of the interface network. For above synthetic polycrystal, though, we can quantify from which grains each ROI obtains its crystallographic signal. For this purpose, we implemented a numerical exact computational geometry method which is detailed in the supplementary material.

The effect of signal mixture close to interfaces, here exemplified for grain boundaries, is best understood for the case of no noise Fig. \ref{fig:verify_px_indexing}b). There are only a few ROIs (blue dots) with $\zeta < 0.5$, i.e. for which the signal comes from at least two grains. Only for these ROIs, the disorientation angle is much higher than \SI{1}{\degree} (the resolution of the grid). The reason that the indexing fails here is because the signatures have a too complex mixture of intensity peaks as it is shown for exemplary cases in Fig. \ref{fig:verify_px_indexing}d). Therefore, indexing with a signature from a single crystal yields an arbitrary match in favour for one of the neighbouring grains if any. The details are dependent on the exact intensity distribution and the individual signal contributions. In effect, the above verification pinpoints that to index reliably at interfaces it needs further work towards e.g. advanced pattern matching or an iteratively refining indexing algorithm --- a situation which is similar to the one for electron diffraction methods \cite{Wright2014,Britton2018}.

We are aware of the fact that microstructures which get instantiated from Poisson-Voronoi tessellations have flat interfaces. To arrive at more realistic interface networks, it is possible to replace the structure synthesis in favour of more advanced protocols from the continuum microstructure modeling community; for instance via interfacing to tools such as DREAM.3D \cite{Groeber2014}. This would add interface facets with different curvature. While causing a different distributing of the ROI volume among the grains, it poses, though, no fundamentally new challenge for the question whether indexing is possible or not.

\subsection{Application to experimental APT specimens}
\paragraph{Aluminium bicrystal}
Finally, we applied the tools on two experimental APT datasets with strong crystallographic information. The first specimen was a technically pure single-phase aluminium bicrystal with a total of $\SI{48.7e6}{}$ ions. The dataset was characterised previously in substantial detail \cite{Wei2019} (R5076-31053-v01). The dataset was crystallographically calibrated, according to settings in the supplementary material. Signatures were processed for several lattice plane families \hkl{002}, \hkl{220}, and \hkl{111}, respectively, for each ROI. The ROIs have a radius of $\SI{20}{\angstrom}$. ROIs were placed on a cubic ROI grid. 

We implemented an iterative approach to refine this grid efficiently to pinpoint at which positions the crystallographic signal is particularly strong. Specifically, a local octree-like grid refinement was implemented. In the first iteration, the dataset was scanned with a coarse cubic ROI grid with ${(\SI{20}{\angstrom})}^3$ spacing. The distribution of signature intensity values for each signature and every ROI is analysed to identify those positions where the maximum intensity per signature exceeds a threshold value (here choosing $\kappa \geq 0.75$ normalised image intensity). In a second iteration, we performed a local refinement of these ROIs by splitting the corresponding ROI grid cell into ${5}^3$ cells and placing that amount of new ROIs at the respective centre of each ROI from the previous iteration. 

We discuss the results for the \hkl{002} signatures in the manuscript while the results for the \hkl{220} and \hkl{111} signatures are included in the supplementary material. Synthetic datasets for aluminium single crystals from the verification were taken as the reference signature. These were also computed specific for \hkl{002}, i.e. composed from the peaks of the same bins.

\begin{figure}[!htb]
\centering
\includegraphics[width=1.0\textwidth]{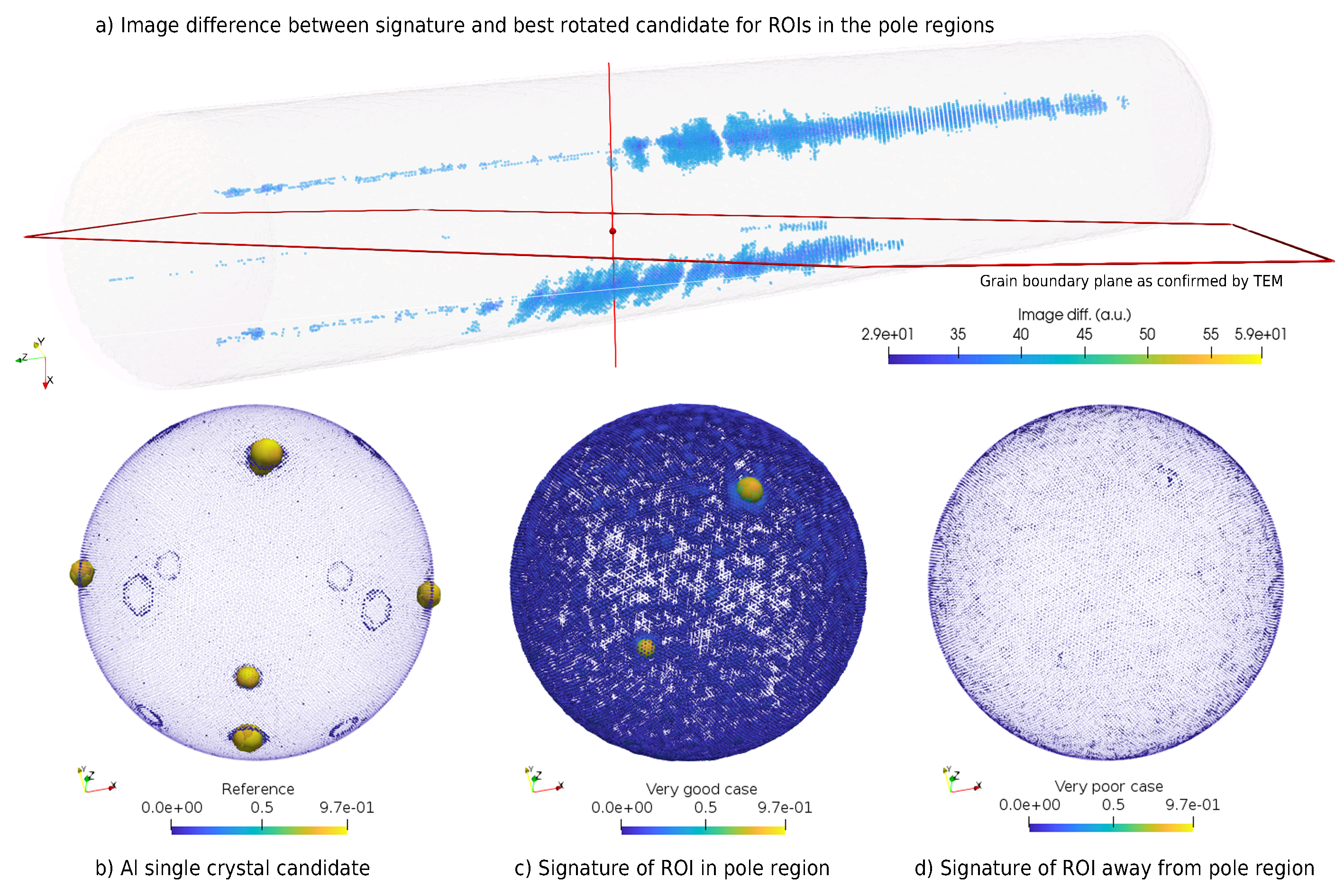}
\caption{Quantification of the crystallographic signal quality within the experimental dataset of the aluminium bicrystal \cite{Wei2019}. Our methods quantify that the crystallographic signal is strongest in the pole regions, Fig. a). The bottom row compare the signature of the reference (a synthetic aluminium single crystal) with the observations for ROIs with a particular good (strong signature) or a particular poor (weak signature) signal quality. Thresholding via the image difference reveals that only for the ROIs along the pole regions (turquoise bluish tube in Fig. a) the signatures are strong enough to pick up crystallographic information. Due to strong lateral distortions, not all of the expected peaks from the theoretical signature appear - even in the good signatures. Therefore, it is possible to back out only information about the possible orientation fibres but not the specific orientation.}
\label{fig:bicrystal_zirong}
\end{figure}

Figure \ref{fig:bicrystal_zirong} shows that for the datasets from real APT experiments of pure aluminium, our method is capable of extracting the crystallographic information content throughout the entire dataset - volumetrically and in an automated manner. Figure \ref{fig:bicrystal_zirong}a) displays a rendering of the reconstructed dataset (grey shading) as well as the grain boundary (confirmed by correlative transmission electron microscopy \cite{Wei2019}). Thresholding was used to visualise those regions in the dataset where our method suggests that the crystallographic information content is highest - here for \hkl{002} in each of the adjacent grains. Given that these pole regions make for approximately only a tenth of the entire dataset volume, our adaptive grid refinement enables to cut numerical costs where the signal is very likely low and invest these computations better at those regions where the signal quality is higher. 

However, the spatial resolution is generally not sufficient to pick up multiple sets of planes within a single ROI as Figs. \ref{fig:bicrystal_zirong}b), c), and d), illustrate.  Figure \ref{fig:bicrystal_zirong}b) shows the reference, the synthetic dataset representing an aluminium single crystal where numerous peaks corresponding to different sets of lattice planes are shown. Fig. \ref{fig:bicrystal_zirong}c) shows an example of a particularly good case from an ROI within the \hkl{002} pole region where the signature has two clear intensity peaks (yellow dots) opposite to each other confirming that two sets of planes have been detectable within this ROI. However, compared to the signature of the reference in Fig. \ref{fig:bicrystal_zirong}b), the four other strong peaks (yellow) are missing in Fig. \ref{fig:bicrystal_zirong}c).

Fig. \ref{fig:bicrystal_zirong}d) also documents that in most cases, the signatures have insufficient signal-to-noise. Virtually no peaks are detectable in these signatures measured outside the pole regions. Pitting against the above verification of the real space method, we can conclude that the reconstruction quality in these regions is either too low to recover the orientation of the crystal or the respective lattice plane set cannot be detected due to geometrical constraints during an APT measurement.

Such cases of missing peaks in the measured signatures makes unique indexing of the orientation based on the information in a single ROI impossible. One would have to analyse at least a second signature for a different set of planes \hkl{hkl} in another pole region to recover the orientation of the grain. While Fig. \ref{fig:bicrystal_zirong}a) clearly shows the dominant \hkl{002} planes have been detected in each grain, other faint signal from other poles can also be observed, combining this information would be sufficient to back out the crystallographic orientation of each grain relative to the detector. 


\paragraph{Al-Li-Mg-Ag alloy}
The second example from experiment generalises the above findings for datasets with second-phase precipitates and covers the situation when these precipitates are small enough to pose challenges with respect to finite counting effects. This is especially the case when the ROI and the precipitate radius is in the order of a few nanometer. The dataset was reconstructed from measuring an Al-Li-Mg-Ag alloy specimen with a dispersion of \althreeli{} $\delta$' precipitates \cite{Villars2016b} (approximate radii \SIrange{22}{39}{\angstrom}) inside a single-crystalline matrix. The specimen was reconstructed and characterised previously in detail \cite{Gault2012e} (R18-15386v01). This particular dataset is a reconstruction from a specimen that was annealed for \SI{8}{\hour} at \SI{150}{\celsius}. We scanned the dataset with a ${(\SI{20}{\angstrom})}^3$-spaced ROI grid ($R = \SI{20}{\angstrom}$). The signatures were compared against single-crystalline references for aluminium, lithium, and \althreeli{}, respectively using peaks specific for \hkl{002}, \hkl{220}, and \hkl{111} lattice plane sets for the individual crystal structures and sub-lattices. Again, we refined the ROI grid once. All results for the grid refinement are available in the supplementary material \cite{Kuehbach2020d}.

\begin{figure}[!htb]
\centering
\includegraphics[width=1.0\textwidth]{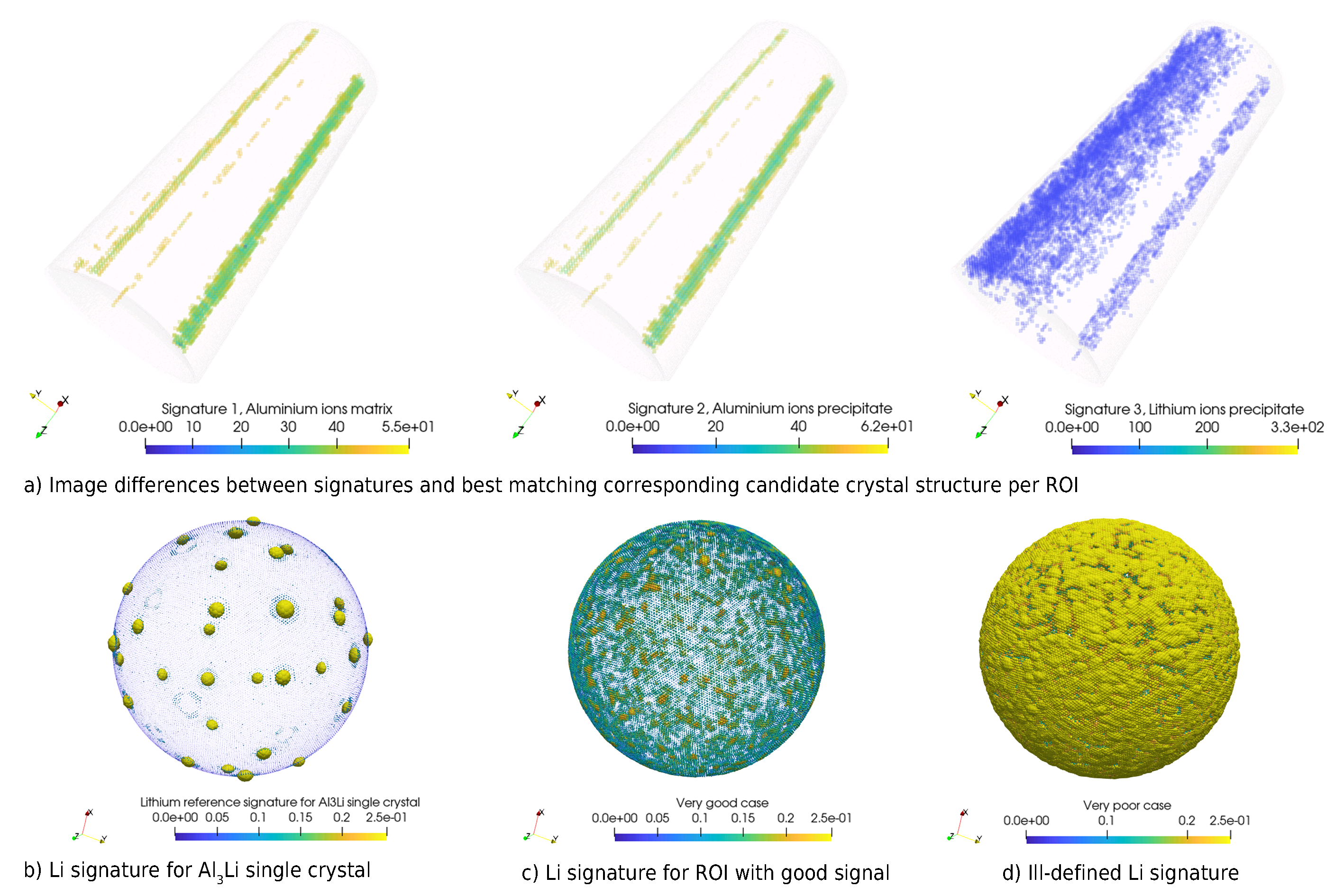}
\caption{Quantification of the crystallographic signal quality within the experimental dataset for the Al-Li-Mg-Ag alloy specimen \cite{Gault2012e}. Thresholding via the image difference of the individual three signatures per ROI highlight again that crystallographic signal is strongest in the pole regions, Fig. a). Figure b) shows the signature of the reference for the ${\textnormal{L1}}_{\textnormal{2}}$ crystal structure - here quantified with the atoms of the lithium sub-lattice. This reference is compared to an exemplar ROI with signatures that were among the strongest of all detected. Last results are compared to an exemplar ROI with ill-defined signatures because of limited lithium counts.}
\label{fig:allimg_baptiste}
\end{figure}

Figure \ref{fig:allimg_baptiste} summarises the main findings, exemplified for a discussion of the \hkl{002} signatures. The automatic approach detects regions in the dataset which have a low image difference between reference and signature, indicating stronger retained signal in the pole regions than elsewhere in the dataset.

An example of finite counting effects is shown in Fig. \ref{fig:allimg_baptiste}d). The example details to which minimum atom count a crystallographic analysis with the real space method can be pushed for a single ROI. The exemplar signature shows strong intensity for virtually all projected directions. An inspection of the individual SDMs confirmed that this was caused by spurious occupation of the spatial distribution map because the ROI contains few lithium atoms. Consequently, the FFT translates such a noisy and eventually even skewed \cite{Haley2019b} histogram into structure. The resulting intensity in the inspected bins of the amplitude spectra is either very low or very high, in the example present here almost close to 1.0 for many ROIs (see the diagram of the signal intensity versus atom count for the Al-Li-Mg dataset in the supplementary material). In effect, indexing becomes an ill-posed task. 

Observation of such finite counting effects, in combination with spatial noise, pinpoints the key difference between algorithmic recovery of structural information from the reconstructed point cloud of a noisy APT dataset versus applying structure identification algorithms on electron diffraction microscopy data: Near atomic resolution is not resolution in the order of thermal lattice vibrations. This concludes our analysis of methods for extracting signatures that capture the long-range periodic arrangement of atoms in 3D point cloud data based on the real space method. With this we resolved a remaining gap in the understanding of the real space method \cite{Araullopeters2015}. Also we deliver a so far missing fully automated method for indexing crystal structure and orientation from noisy point cloud data in experimental datasets. It was suggested recently \cite{Haley2019b} that this is of interest for APT.

By developing a more performant set of tools and strategies for rigorously testing these, we opened the door for a more productive high-throughput workflow within atom probe crystallography. As a combination of open-source \python{} scripts and compiled scientific computing tools, our work can facilitate experimentalists in knowing which regions contain crystallographic signal and also quantify the relative signal strength of this information. This helps to supplement atom probe crystallography studies which characterise poles in detector space or extract interfaces between crystals via elemental segregation. To conclude, such uncertainty quantification can help to improve open data exchange in the quest for improving reconstruction algorithms.

\subsection{Benchmarking}
\begin{figure}[!htb]
\centering
\includegraphics[width=0.5\textwidth]{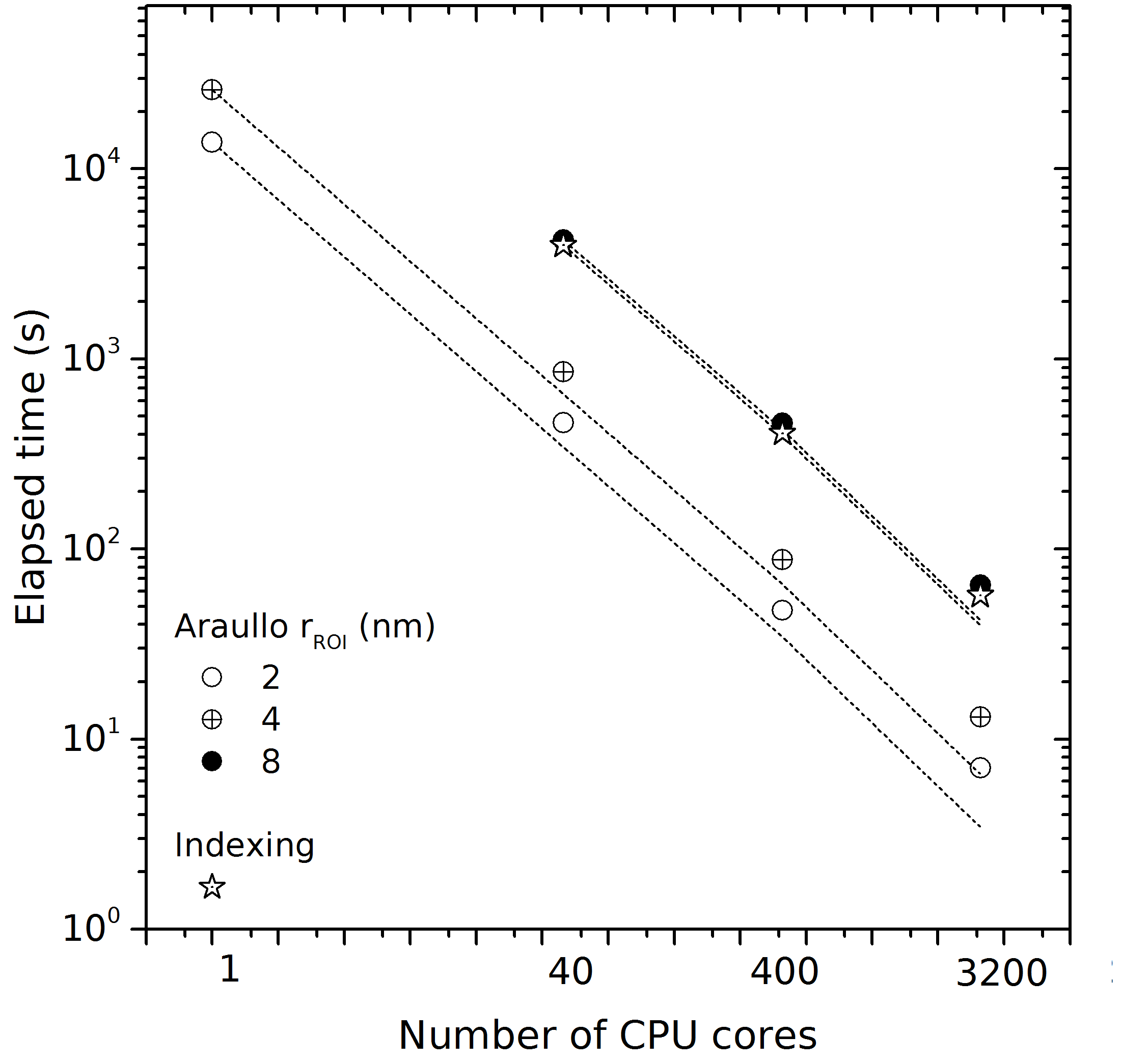}
\caption{Elapsed time results of the same setups processed with an increasing number of CPU cores document the strong scalability of the \protect{\araullo{}} tool. Acquisition with \SI{80}{\angstrom} (black circle symbols) took approximately the same time as indexing (open star symbols). The straight dashed lines compare the results to the theoretical case of ideal linear scaling. These benchmarks processed always \SI{1e4}{} ROIs for the synthetic aluminium single crystals. Different ROI radii \SIrange{20}{80}{\angstrom} are compared for the same binning \ensuremath{m = 10}.}
\label{fig:benchmark_araullo}
\end{figure}

\begin{figure}[!htb]
\centering
\includegraphics[width=0.5\textwidth]{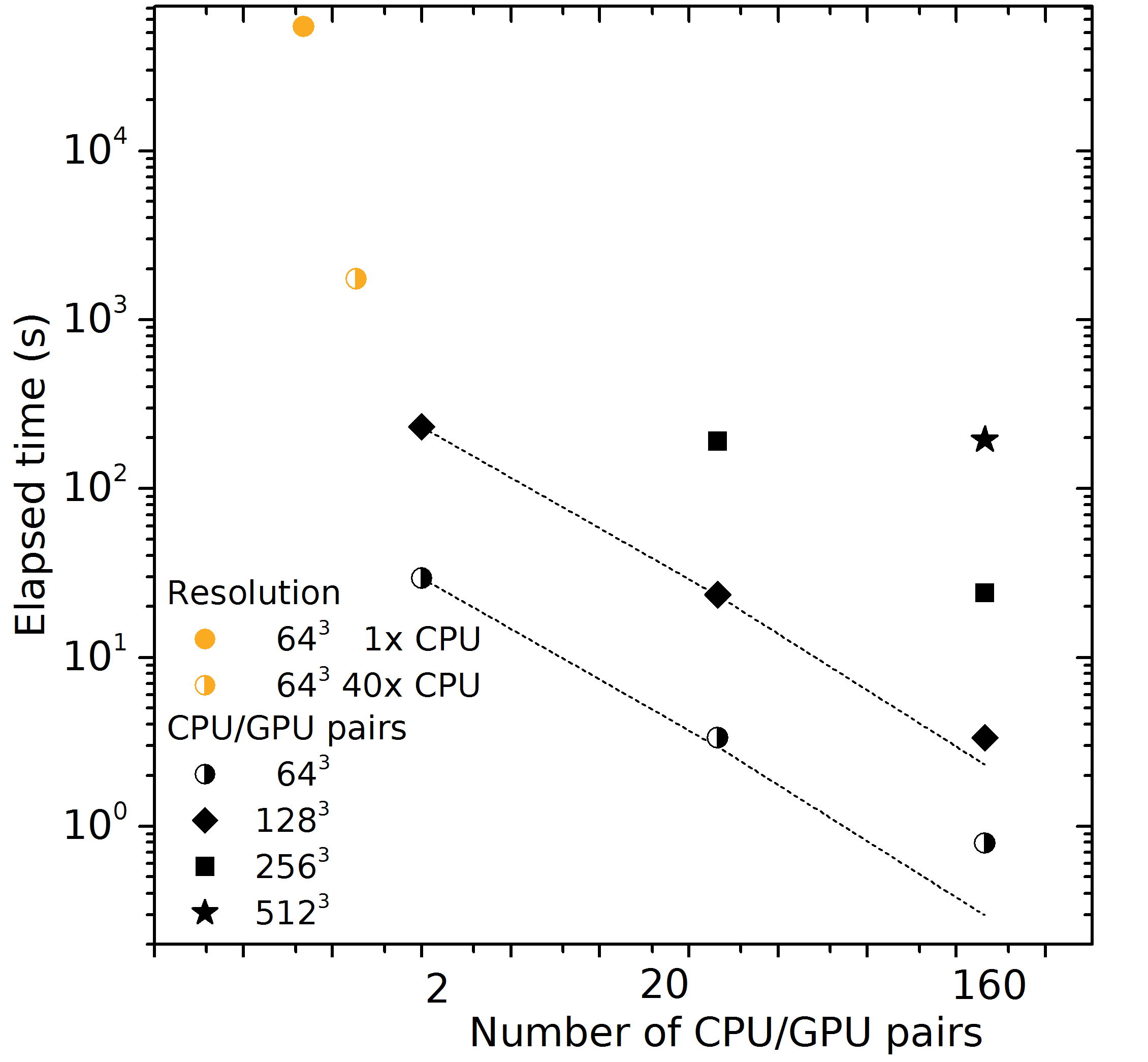}
\caption{Elapsed time results of the same setups processed with an increasing number of CPU/GPU pairs document the strong scalability of the \protect{\fourier{}} tool. Different reciprocal-space resolutions ($64^3$ - $512^3$) were tested. Straight dashed lines again compare to linear scaling. The benchmarks probed \SI{1e4}{} ROIs with \ensuremath{R = \SI{20}{\angstrom}} and the same synthetic aluminium single crystals.}
\label{fig:benchmark_fourier}
\end{figure}

\paragraph{Real space method}
For practitioners it remains to document the performance and scalability of the tools. To quantify the strong scalability \cite{Kuehbach2020a}, we executed above case studies with an increasing number of CPU cores or GPUs. The results are shown in Figs. \ref{fig:benchmark_araullo} and \ref{fig:benchmark_fourier}, respectively. The dashed lines are linear extrapolations of the elapsed time reduction under the assumption that adding more CPU cores or GPUs results in a proportional reduction of the elapsed time \cite{Amdahl1967}.

If executed sequentially (one CPU core), the benchmark with the real space method took \swisswatch{3}{50} for $R = \SI{20}{\angstrom}$ and solving $\SI{1e4}{}$ ROIs. Using $40$ CPU cores of a single node brings down the elapsed time to less than \SI{8}{\minute}. This is a strong-scaling multithreading efficiency of at most $\SI{76}{\percent}$. Using more cores results in higher productivity. As an example, solving the same benchmark above with $3200$ CPU cores takes $\SI{13}{\second}$ elapsed time. This is an approximately 2000-fold performance increase or $\SI{63}{\percent}$ strong-scaling efficiency. Two studies report performance data for the real space method \cite{Araullopeters2015,Haley2019b}, however they support only sequential execution. The results are difficult to compare with the present findings because different soft- and hardware as well as settings were used. It seems though, that these two studies are at least sequentially in the same order of performance. By contrast, our approach is scalable.

\paragraph{Reciprocal space method}
The reciprocal space method was executed with CPU multithreading and alternatively with GPU processing (Fig. \ref{fig:benchmark_fourier}). For a set of $\SI{1e4}{}$ ROIs with a radius of $R = \SI{20}{\angstrom}$ and resolving the reciprocal space with a ${({\mathcal{L}}_D = 64)}^3$ grid, the computation takes \swisswatch{15}{4}. With $40$ cores, results are ready after $\SI{29}{\minute}$, thereby documenting a strong-scaling efficiency of $\SI{78}{\percent}$. Using already a single GPU, though, outperforms the $40$ CPU cores by well an order of magnitude. Eventually using $160$ GPUs of the TALOS cluster enabled to complete the benchmark in $\SI{0.8}{\second}$ for a reciprocal space grid with ${({\mathcal{L}}_D = 64)}^3$ points and $\SI{200}{\second}$ for ${({\mathcal{L}}_D = 512)}^3$ points, respectively. Tapping such, so far unused resources, enables hitherto inaccessible uncertainty quantifying for APT data.

\section{Discussion}
The work is not complete without bridging to existent work on indexing crystal structure and orientation developed by other (microscopy) communities. Building these bridges is the purpose of the last section to envision possible strategies for inspiring either development in these fields or improve the above methods in the future.

\subsection{Spherical harmonics methods for representing the signatures and indexing them?}
Indexing signatures against a library of precomputed rotated references is a brute force approach because it demands for each ROI the processing of a large set of rotations and large number of image intensities. Instead, it might be more efficient to use alternative methods like spherical harmonics for solving this image registration task \cite{Makadia2006a,Makadia2006b} and blend this with spectral methods that were proposed recently in the (X-ray diffraction) texture and SEM/EBSD community \cite{Hielscher2019,Lenthe2019}. 

The key idea reads as follows: an image, carrying the crystallographic information, is compressed into a reduced-order description. Next, it is projected in such a way into a mathematical space that particular mathematical rules can be applied to solve the registration \textit{and} orientation task more efficiently. Assessing and comparing such a method to ours is worth its own careful analysis. Therefore, we explore in this paper only whether it is possible to create such reduced-order description of the signatures.

As a key requirement, a series expansion may be used to fit the intensities of the signature. The weights, or coefficients, of the series expansion afford a reduced-order description of the spherical image. The values of the series expansion are intended to vary sufficiently in order to fit to a generalised distribution across the surface of a sphere. This is an application for discrete spherical harmonics. 

\begin{figure}[!htb]
\centering
\includegraphics[width=0.75\textwidth]{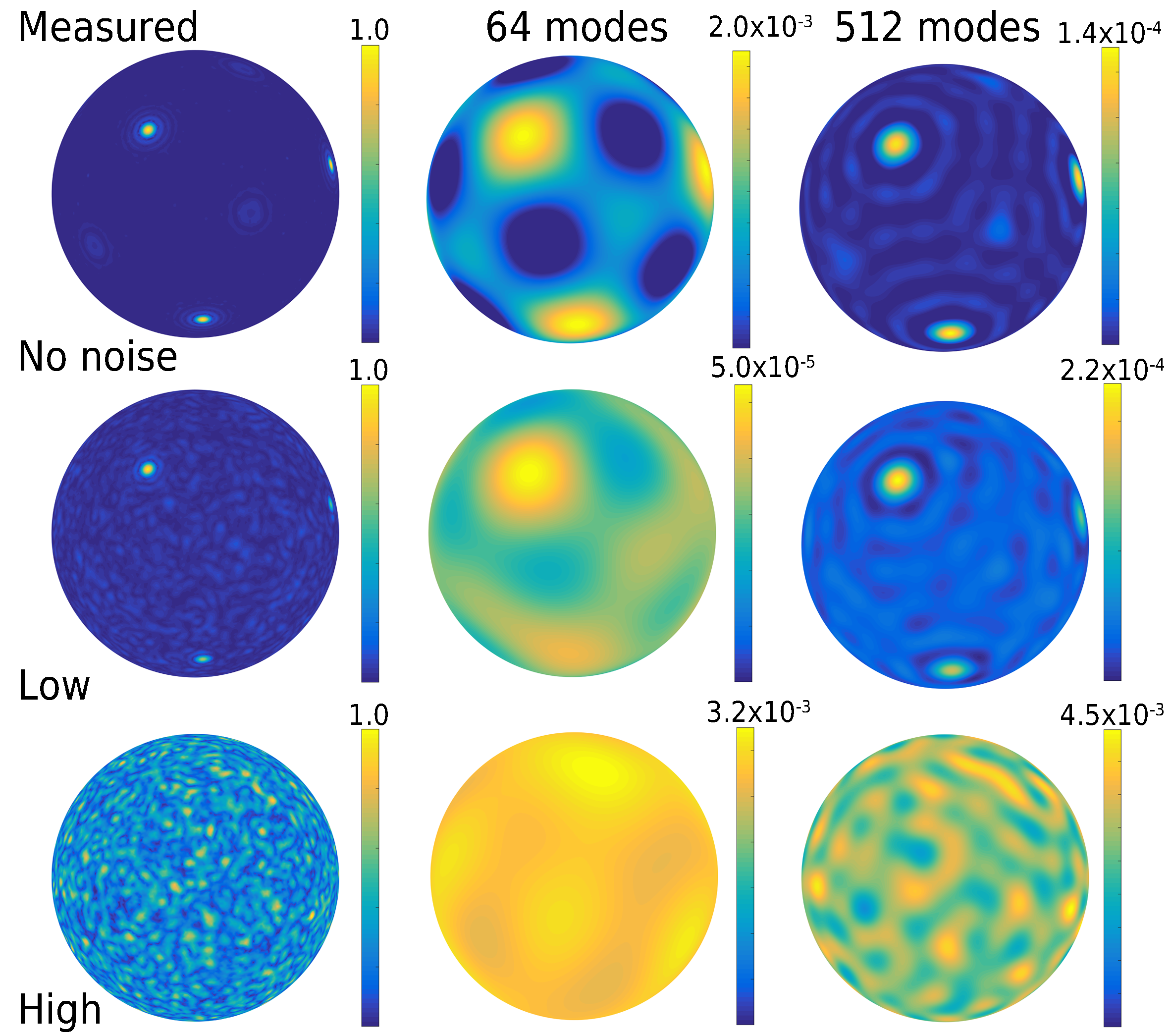}
\caption{Preliminary results on approximating the signatures for ROIs probed in the synthetic aluminium single crystal for different strength of positional noise. The colour bars display the image intensity using individually a linear scale with values that range from $0.0$ to the shown individual maximum intensities. Spherical harmonics approximated spherical images (of the signatures in the left column). We compared approximation for $64$ and $512$ spherical harmonics modes, respectively.}
\label{fig:matthew_sph}
\end{figure}

Here, we employ a finite element approach and evaluate sequential \matlab{} code to find the corresponding weights of the discrete harmonic series expansion. Specifically, we adapt a strategy for fitting spherical harmonics to lattice strain pole figure intensities \cite{Wielewski2017}. The key mathematics are recapped in the supplementary material because we only replace the pole figure intensities by image intensities of the signature. As an example we inspected spherical harmonics descriptions of several signatures from the synthetic aluminium single crystals in \bunge{8.0}{8.0}{8.0} orientation, using the single crystal reported earlier from the method verification. The results are summarised in Fig. \ref{fig:matthew_sph}.

First, we focus on the image matrix in Fig. \ref{fig:matthew_sph} to check which information content of the signatures the spherical harmonics recover. The rows compare the signatures (left column) with the fitted signatures (right column). We compare for increasingly stronger positional noise - a perfect crystal ($\sigma_z = \SI{0.00}{\angstrom}$, left column, top row), a crystal with low noise ($\sigma_z = \SI{0.25}{\angstrom}$, left column, middle row), and a crystal with high noise ($\sigma_z = \SI{0.75}{\angstrom}$, left column, bottom row), respectively. The columns of the matrix compare these signatures with the approximation of the intensities via spherical harmonics using $64$ or $512$ harmonic modes, respectively. The spherical harmonics pick up the location of the intensity peaks already with a low number of modes. However, this comes at the cost of substantially smeared out intensities and moderate approximation quality improvement for an increasing number of modes. 

In turn, our explored combination of a fine FE mesh and orientation grid resolution resulted in angular accuracy and precision very close to the best so far achieved $\SI{0.5}{\degree}$, as it was shown experimentally in correlative TKD microscopy and APT studies of Breen \textit{et al.} \cite{Breen2017}. Consequently, we wish that a proposed spherical harmonics algorithm should ideally be equally accurate and precise. In this regard, the results in Fig. \ref{fig:matthew_sph} are preliminary. Nevertheless, they suggest that spherical harmonics could be a useful alternative for compressing signatures. 

\subsection{Methods from TEM as an alternative to index APT data?}
We observe that the crystallographic images from the reciprocal space method are almost equal in format to Automated Diffraction Tomography (ADT) datasets from TEM diffraction studies. To this end, algorithms have been developed in the TEM community \cite{Campbell1998,Kolb2006,Kolb2008,Kolb2011,Maia2011} which back out the crystallographic orientation for a known crystal structure. Eventually, these algorithms are even capable of detecting the most likely crystal structure candidate. Observing the recent progress in the field of electron nanodiffraction \cite{Zuo2019}, this could be an option to explore in another interdisciplinary atom probe crystallography study in the future.

\subsection{Artificial intelligence methods as an alternative to index APT data?}
Finally, it is worth mentioning that similarities exists between the reciprocal space method and recently proposed deep-learning approaches for identifying crystal structures \cite{Ziletti2018,Leitherer2019a}. Similar is that both methods encode the structural information via an image of (direct) Fourier-transformed atom positions. Another similarity is the formulation of indexing as an image processing task. 
 

Neither of the above methods nor the artificial intelligence methods work without computing a signature. In the artificial intelligence approach, these are the raw data and respective inputs for training and inference. It can be seen as a benefit of deep-learning approaches that the features of the signatures do not need to be encoded manually because they are evaluated as part of the feature mapping during training and inference. By contrast, in our indexing approach with the real space method, it is necessary to define \textit{a priori} from which peak(s) in the amplitude spectra the signature is composed. We have shown that this is possible for some crystal structures, provided there are no ambiguities. Resolving these can be more complicated, though, for arbitrary crystal structures, especially those with a low symmetry. 

A clear disadvantage of above deep-learning approach is that it does so far not account for the relative orientation of the crystal. This holds at least as long as a descriptor like the smooth overlap of atomic potentials (SOAP) \cite{Bartok2013,De2016} is employed, which is typically formulated rotation-invariant. At least technically such limitation could be lifted, though, and combined with a training on explicit sets of candidate orientations for each crystal structure via data augmentation. 

Alternatively, deep-learning could be combined with above-mentioned spherical harmonics description. Blending our methods with artificial intelligence could help to improve in cases where discussed signal mixtures at interface junctions are challenging. Nevertheless, it remains to be shown that the robustness of above deep-learning methods holds also for noise levels as high as they were inspected in this work.

\section{Summary}
We developed a method for extracting crystal structure and orientation volumetrically within reconstructed atom probe tomography datasets. The tool works accurately and precisely. It is fully automated and scales strongly on parallel computers. The methods are delivered as open-source software to contribute to faster and more reliable atom probe crystallography. Verification and validation studies with synthetic datasets and experimental APT specimens brought the following conclusions:

\begin{itemize}
    \item Suitably modified, the method of Araullo-Peters \textit{et al.} yields reproducible and robust crystallographic information for spatial uncorrelated missing atoms or datasets where the atoms are as much as \SI{10}{\percent} of the lattice spacing displaced out of their equilibrium positions.
    \item Using a different data post-processing strategy, the method becomes a tool for identifying crystal structure and orientation. We verified and validated its functionality for noisy datasets from single crystals and polycrystals with an arbitrary grain boundary network.
    \item Software parallelisation enables now analyses on computer clusters, using either CPUs alone or a combination of CPUs and GPUs. Benchmarks with at most \SI{3200}{} CPU cores or \SI{160}{} GPUs, respectively, delivered three orders of magnitude faster processing compared to previously reported sequential tools.
    \item These achievements can assist the development of more precise reconstruction methods by enabling fast and reliable quantitative assessments of the crystallographic information content in reconstructed APT datasets.
    \item A quantitative assessment of material points in close proximity to grain boundaries and triple lines addressed the deterioration of the crystallographic information due to signal contributions from multiple grains. Strategies for  improvements were sketched and discussed in relation to alternative methods from crystal structure identification of the electron and X-ray diffraction communities.
    
    \item The automated method was tested on an experimental aluminium bicrystal and an Al-Li-Mg-Ag alloy. Crystallographic signal was detected in the poles regions. In some cases, two sets of planes could be detected in a single ROI but in most cases only one set of planes could be detected in each pole. In regions outside poles, the method did not detect resolvable structure. Nevertheless, the method enables a fast and powerful way to detect and quantify latent crystallographic information within experimental APT reconstructions in 3D that until now has had practical limitations due to high computational complexity which we mitigate with highly performant parallelised and CPU- and GPU-optimised code. 
\end{itemize}

\section{Data and source code availability}
The source code of the tools \cite{Kuehbach2020b}, hands-on tutorials in the form of \python{}/Jupyter notebooks for practitioners \cite{Kuehbach2020b,Kuehbach2020c}, as well as the configuration files and datasets for software developers \cite{Kuehbach2020d} are available as open-source supplementary material.

\section*{Work distribution}
MK\"U designed the study, implemented \paraprobe{}, and led the manuscript writing. MKA contributed the geodesic finite element mesh and \matlab{} scripts for fitting signal intensities using spherical harmonics. ABR contributed the in-depth crystallographic analyses of the experimental results. All authors discussed the results and contributed to writing of the manuscript.

\section*{Acknowledgements}
MK\"U and ABR acknowledge the financial support by \textit{BiGmax, the Max Planck Society's Research Network on Big-Data-Driven Materials-Science}. ABR is grateful for funding from the \textit{Alexander von Humboldt Foundation}.

\newpage
\section*{Supplementary material}
For the purpose of this preprint we have added the pages that we plan as supplementary material below.

\section{Computational methods}
\subsection{Setting up atom probe crystallography workflows with \paraprobe{}}
\paragraph{Input and prerequisities}
Setting up a workflow to the analyses in the main paper starts with defining four ingredients: a dataset (synthetically created or reconstructed from APT experiments), a ranging, i.e. mapping of the mass-to-charge ratios to atom types, an ensemble of ROIs, and a set of crystal structure candidates with individual definitions along which crystallographic directions one wants to probe, and hence which region(s) of the amplitude spectra to investigate. We worked with synthetic datasets as well as datasets from real APT experiments. Synthetic datasets were created using the \synthetic{} tool \cite{Kuehbach2020a}. The synthetic datasets were on the one hand used to build the datasets for verifying the tools and on the other hand used to compute the reference signatures for single crystals with specific crystal structure and orientation.

For the experiments we imported reconstructed atom positions by transcoding either POS or EPOS files, respectively from \ivasfour{} \cite{Ulfig2017,Reinhard2019} into the Hierarchical Data Format (HDF5) file representation required for \paraprobe. For this we worked with the \transcoder{} tool. For ranging, \paraprobe{} accepts ranging data and definitions that have been made with external tools from the APT community. Such ranging data define how to map from mass-to-charge ratios to atom types. Specifically, we worked with the import functionalities of the \ranger{} tool and read RRNG files. The ranging was created by the respective APT experimentalists within \ivasfour{}. The definition of the ROI ensemble and the crystal structure candidates is detailed in the main paper for each respective case study.

\paragraph{Workflow}
Quantifying crystallographic signal and indexing it via \paraprobe{} is realised with a sequential workflow of \paraprobe{} tools. Individually, these tools use parallelisation. The steps for the real space method were as follows:

\begin{enumerate}
    \item We synthesise synthetic datasets for the crystal structure candidates, range these, and characterise their atomic architecture. Thereafter, we evaluate the amplitude spectra for each spatial distribution map (SDM) and decide which lattice plane stack to analyze. Based on this we decide which region of the amplitude spectra we need to probe and compose into signatures:
    \begin{enumerate}
        \item \synthetic{},
        \item \ranger{},
        \item \surfacer{},
        \item \araullo{}.
    \end{enumerate}
    \item The analysis for the dataset to be indexed uses above created reference signatures for the crystal structure candidates:
    \begin{enumerate}
        \item \transcoder{} or \synthetic{}, respectively depending on whether data come from experiment or are synthetic,
        \item \ranger{} to accept the external ranging for the dataset,
        \item \surfacer{} to compute the distance of the ROI to the dataset edge,
        \item \spatstat{} to optionally characterise spatial statistics,
        \item \araullo{} to compute the signatures for all ROIs,
        \item \indexer{} to index the signatures against the rotated references for the above defined (crystal structure) candidates.
    \end{enumerate}
\end{enumerate}

\begin{figure}[!htb]
\centering
\includegraphics[width=\textwidth]{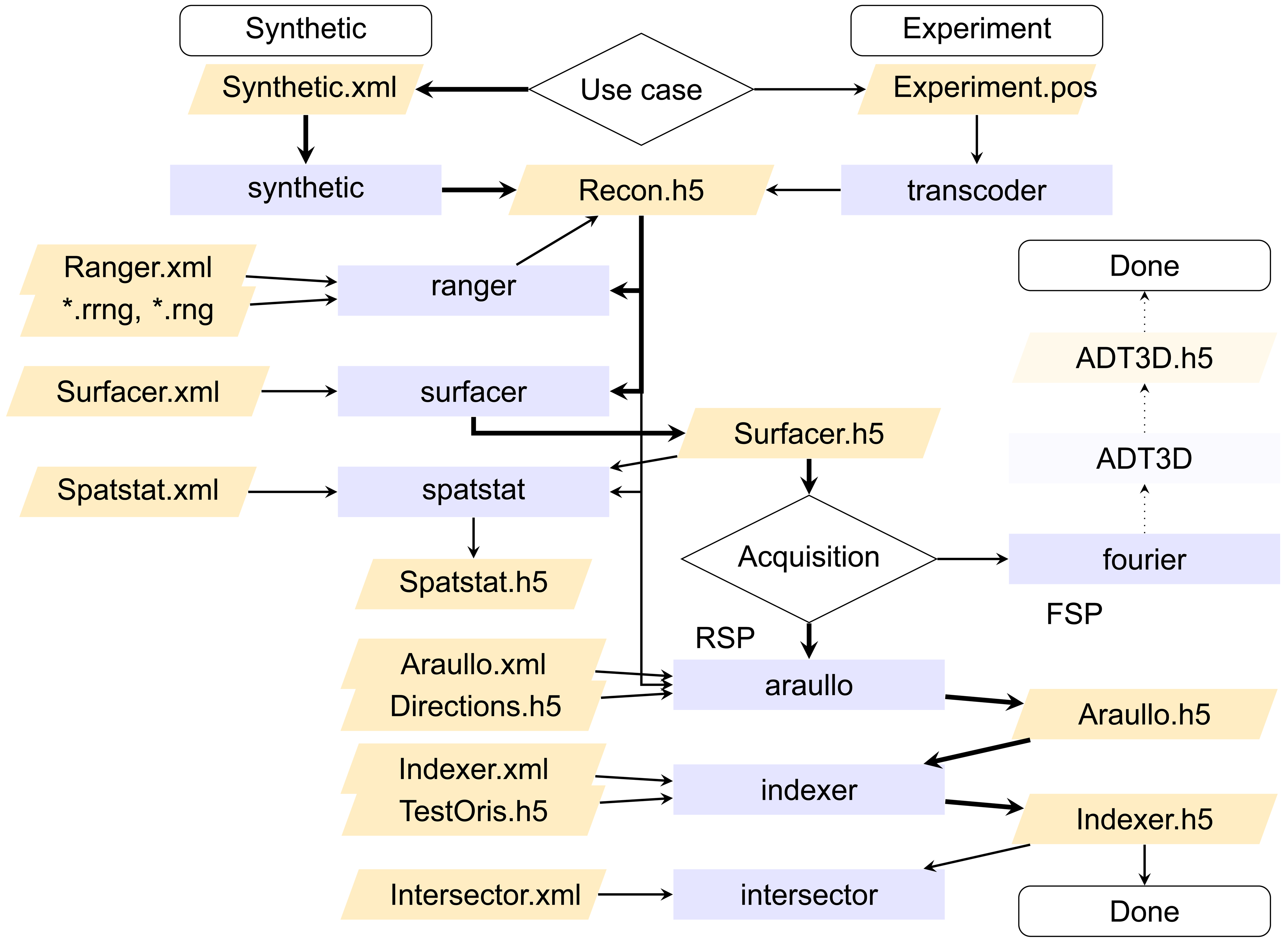}
\caption{Proposal for an automated workflow for indexing crystal structure and backing out crystal orientations locally. Thicker arrows follow the main steps of the workflow from the creation of a synthetic or the importing of a reconstructed dataset from an experiment to the acquisition of the signatures. RSP abbreviates the real space method. FSP abbreviates the reciprocal space method. Instead of a monolithic software, we connect multiple specialised tools (blue boxes) into a workflow. The analysis steps are controlled through XML configuration files. All heavy data are carried through the processing with HDF5 files.}
\label{fig:key_concepts}
\end{figure}

Figure \ref{fig:key_concepts} summarises the resulting workflow and how the real space and reciprocal space methods, respectively are integrated into this workflow.

The GitLab repository of \paraprobe{} \cite{Kuehbach2020b} comes with a set of Jupyter notebook tutorials which exemplified further details of above workflow.

\subsection{Implementation details}
\paragraph{Assuring that ROIs are completely in the dataset}
To assure this, we first binned each dataset into $\SI{10}{\angstrom}$ cubic voxels. Then, we identified the interior voxels using methods previously described \cite{Kuehbach2020a}. Regions of interest (ROIs) were placed only in such voxels. In addition, dense \ashapes{} of the point cloud were computed \cite{Kuehbach2020a}. The resulting triangle hulls were evaluated for intersection with the ROIs. Only ROIs without intersections with the \ashapes{} were used. In combination, this ensures that the ROIs are fully embedded in the dataset.

\paragraph{Why geodesic spheres instead of a regular gridding of elevation/azimuth?}
Multiple strategies exists to define along which projection directions one probes for the real space method. Compared to a regular gridding of the elevation-azimuth space, as it was proposed in \cite{Araullopeters2012}, we work with a geodesic sphere finite element mesh because it distributes the nodes more homogeneously over the surface of the unit sphere \cite{Popko2012}.

\paragraph{Controlled windowing of the SDMs}
We used a different binning strategy than the original authors \cite{Araullopeters2012} for preparing the SDMs before the FFT was computed. Specifically, we pad the histogram on either side by one bin with zero counts to enable for any implementation of controlled signal windowing \cite{Prabhu2013}. In this work, we used a rectangular windowing function. Alternatively, Kaiser windowing could be used \cite{Kaiser1980}.

\paragraph{Creation of test orientations for rotating the reference signatures}
We worked with a quasi-equidistant grid of test orientations in orientation space $\mathcal{G}$ with members $g \in \mathcal{G} \in SO3$ using the \matlab{} \mtex{} texture toolbox (v5.0.3) \cite{Bachmann2010b,Bachmann2010}. Crystal symmetries were accounted for ($\frac{4}{m}\bar{3}\frac{2}{m}$). The angular resolution of the grid was $\SI{1}{\degree}$, resulting in a total of $N_g = 618324$ orientations. The rotations are evaluated in a pre-processing stage to build a library of rotated references (for each crystal structure candidate). Orientation differences were quantified using classical disorientation-based algorithms \cite{Grimmer1974,Heinz1991}. Also these were implemented for cubic crystal symmetry only, generalisations have been reported \cite{Bonnet1980,Heinz1991,Morawiec2004}. The disorientation quaternion in this work is defined as a particular misorientation quaternion in the fundamental zone with the smallest rotation angle.

\paragraph{Indexing algorithm}
Running the real space method for an ensemble of ROIs yields a collection of spherical images, i.e. the signatures ${\mathcal{S}}^{msr}_c$ for each $c$-th crystal structure candidate. During indexing we evaluate nodal intensities using the \indexer{} tool (Fig. \ref{fig:key_concepts}). Indexing has three pre-processing and one indexing step. First, the intensities of the ${\mathcal{S}}^{msr}_c$ signatures and the rotated references \textbf{${\mathcal{S}}^{g}_c$ } (signatures for the crystal structure candidates rotated by $g$) are normalised individually. The index $g$ denotes the test orientation. Second, a user-defined sub-set of the FE nodes is defined (here $\hat{N_v} = 1000$) for each reference. We pick those nodes for which the signal is strongest. Third, orientation set $\mathcal{G}$ is used to rotate each of the $\hat{N_v}$ FE nodes and compute its closest neighbouring node $p^g_{v,c}$.  The index $v$ denotes a particular of the strongest intensities in decreasing order. 

The result of the pre-processing steps is a look-up table. This look-up table encodes implicitly which nodal values of the signatures for the ROIs, i.e. ${\mathcal{S}}^{msr}_c$, have to be inspected to compare them with a specifically oriented version of a rotated reference. In effect, indexing reduces to a querying and comparing of image intensity values at pre-computed nodes. The difference $\Delta^g_c$ is quantified as the sum of squared image intensity differences (Eq. \ref{eq:araullo_solution_quality}). For each rotated reference in orientation $g$ and each crystal structure candidate $c$ we get one value for each ROI.

\begin{equation}
\Delta^g_c = \sum^{\hat{N_v}}_{v=1} {({{{\mathcal{S}}^{msr}_c}}_{|_{{p^g_{v,c}}}} - {{{\mathcal{S}}^g_c}}_{|_{v}})}^2 
\label{eq:araullo_solution_quality}
\end{equation}

\paragraph{Analytical intersection volume}
We state in the main paper to have developed a numerical exact method for computing the intersection between a spherical ROI and an arbitrarily shaped and oriented polyhedron. In this work, the polyhedra are the Poisson-Voronoi cells from the polycrystal construction that we interpret as the grains of the polycrystal. To the best of our knowledge there is no analytical formula to compute such intersection volume for arbitrary spatial arrangement of the geometric primitives. 

However, there is a numerical method for computing the exact intersection volume of a sphere cutting an arbitrary tetrahedron \cite{Strobl2018}. Recall, that one can decompose a convex polyhedron into a set of tetrahedra. In our case, we can thus decompose the volume of each Voronoi cell completely into a set of tetrahedra. These tetrahedra are boundary-conformant with the piece-wise linear complex that is represented by the faces of the Voronoi cell \cite{Si2015}. In effect, this enables a computation of the intersection volume between the ROI and each tetrahedron. This allows us to accumulate the individual values for ROI-sphere-tetrahedra-intersections to obtain a total intersection volume between the ROI and each Voronoi cell, i.e. grain $k$.

\paragraph{Sequential implementation tricks}
The numerical costs for computing fast Fourier transforms of the real space method are reducible with numerical libraries such as the \imkl{} \cite{Intel2019} or \fftw{} \cite{FFTW2005} for the CPU, and \cufft{} \cite{Nvidia2019} for the GPU, respectively. Conjugate-even symmetry was assumed to reduce further the costs of the Fourier transformations. Single precision was used where possible. All data were packed into contiguous pieces of memory to reduce memory traffic and cache misses. Although we have not explicitly implemented it, one could also exploit that all histograms (SDMs) for a ROI have the same number of bins. In principle, this enables a batching of all the Fourier transformations for a ROI. Thereby, one can further reduce the numerical costs for loading intermediate values. We queried atoms for each ROI OpenMP-multithreaded.

\subsection{Analysis of asymptotic computational costs}
\paragraph{Real space method}
The numerical costs of the real space method are defined by four quantities: the number of ROIs $N_{roi}$ in the ensemble, their radius $R$, the number of SDMs per ROI $N_v$, and the number of bins per SDM $N_{b} = 2^m$. For a dataset of volume $V$ with homogeneous density of a single atom type $\rho$, and $f$ being a geometrical constant, the average number of atoms in a ROI is $N_w = fR^3\rho$. Therefore, the asymptotic computational time complexity of collecting crystallographic signal with the real space method is:
\begin{equation}
    \mathcal{O}_{Proj} = N_{roi} \cdot [fR^3\rho \cdot log(V\rho) + N_v [N_w + 0.5N_b log(0.5N_b) + 0.5N_b]].
\end{equation}

The summation terms account for the querying, the projecting, the Fourier transforming, and the peak identifying from the amplitude spectrum. Here we assume an algorithm with tree-based neighbour querying \cite{Kuehbach2020a} and amplitude spectra analysed within a selected frequency interval. Despite the numerical complexity, there is substantial potential for parallelising those computations. In fact multiple computations are independent: each ROI, each direction, and the peak search for each amplitude spectrum. Our implementation is the first which taps this potential.

\paragraph{Reciprocal space method}
For the reciprocal space method we probed cubic sub-regions in reciprocal space at reciprocal space positions $\bm{k}$. Different resolutions were probed (${\mathcal{L}}_D = $ \SIrange{64}{512}{} for $\bm{k} \in {[-2\pi, +2\pi]}^3$). The asymptotic computational time complexity of the reciprocal method is:

\begin{equation}
    \mathcal{O}(N_{roi} \cdot N_w \cdot {{\mathcal{L}}_D}^3),
\end{equation}

where $N_w$ is the number of atoms in the ROI. Given the resources available in 2001 \cite{Vurpillot2001}, this rendered routine application of the method impractical. However, the computations are independent for every ROI and have low summation costs for every atom $w$ and every reciprocal space position $k$. In combination with the CPU and especially GPU hardware improvements in the last 20 years \cite{Hennessy2012,Rauber2013,Reinders2014,Jeffers2015}, this makes the proposal and method of Vurpillot \textit{et al.} potentially attractive again for quantifying crystallographic signal.

\paragraph{Indexing}
Our implemented indexing method has a computational time complexity with two key contributions: a constant look-up table creation cost ${\mathcal{O}}_{lu}$ and indexing costs ${\mathcal{O}}_{idx}$. The creation costs are ${\mathcal{O}}_{lu} = N_c \cdot N_g \cdot \hat{N_v}$, in which $\hat{N_v}$ is the number of FE nodes at which image intensities are evaluated and $N_c$ the number of crystal structure candidates. The indexing costs are ${\mathcal{O}}_{idx} = N_{roi} \cdot N_c \cdot N_g \cdot \hat{N_v}$. Both contributions offer substantial potential for parallel execution. The computation of a look-up entry is an independent task per entry. Equally, each ROI is independent and at this level also each orientation and phase candidate. 


\subsection{Soft- and hardware details}
These methods were implemented as additional tools (\protect\araullo{}, \protect\fourier{}, \protect\indexer{}, \protect\intersector{}) into the open-source \protect\paraprobe{} toolbox. The tools were compiled with the Intel (v18.0.5.20180823) compiler using Skylake CPU-specific optimisation. Algorithms from the Computational Geometry Algorithms Library CGAL \cite{CGAL2018,Da2018,Kuehbach2020a} (v4.11.3) and Boost C++ \cite{Schaeling2014} (v1.66) were employed. When using the GPUs, code was compiled with the PGI (v19.10-0 LLVM) compiler \cite{PGI2020} using Skylake CPU-specific and GPU-specific optimisation for the NVidia Tesla GPU architecture. OpenACC and CUDA commands were compiled using CUDA v10.0.130. All tools were linked against the Intel MPI library (v2018.4). I/O operations were executed via the sequential HDF5 library (v1.10.2) \cite{Prabhat2014,HDF52020}.

The analyses were executed on the TALOS computing cluster, a SUSE Linux Enterprise Server 15.1 SP1 system with 80 accessible computing nodes. Each node has two Intel Xeon Gold 6138 twenty-core processors with access to \SI{188}{\giga\byte} main memory in total. Furthermore, each node is equipped with two \vtesla{} \cite{Nvidia2017} accelerator cards with \SI{32}{\giga\byte} memory each. 

The Message Passing Interface (MPI) library \cite{Gropp1998,Gropp1999a,Gropp1999b} was used to distribute ROIs across computing nodes at the coarse scale. At the finer scale, ROIs were delegated to Open Multi-Processing (OpenMP) \cite{Chapman2007} threads. The OpenMP threads were pinned to specific cores using OMP\_PLACES=cores and executing one MPI process per node. In cases were the tools used the CPUs and GPUs, both GPUs of each node were utilised. Each GPU was instructed by one MPI process. Each MPI process spawned for this one OpenMP thread. All resources were used exclusively. The figures were generated using Paraview \cite{Ayachit2015} and Python. Analyses employed single precision, except for TetGen where double precision was employed. 

Explicit calls to the MPI\_Wtime and omp\_get\_wtime functions were made to monitor how much elapsed time the individual workflow steps took. I/O and non-I/O operations were distinguished. Two system variables were parsed on-the-fly to quantify virtual and resident main memory set sizes. These pieces of information were parsed from the /proc/self/stat system file.

\subsection{Reconstruction of the experimental datasets}
\paragraph{Aluminium bicrystal}
We used the following reconstruction parameters:
\begin{itemize}
    \item Image compression factor $\textnormal{ICF} = \SI{1.53}{}$,
    \item Field factor $k_f = \SI{4.91}{}$,
    \item Flight path length ${\mathcal{L}}_{fp} = \SI{100}{\milli\meter}$,
    \item Detection efficiency $\eta = \SI{0.8}{}$.
\end{itemize}

For the real space method we computed $N_v = 40962$ SDMs per ROI with $m = 8$ binning. The signatures ${\mathcal{S}}^{msr}_c$ were composed by assigning each node the highest amplitude of the its corresponding amplitude spectrum in the bin interval corresponding to $\pm$\SIrange{1.823}{2.230}{\angstrom}. Conceptually, this is equivalent to an indexing where one probes for all six crystallographic directions \hkl<100>. In total $\hat{N_v} = 1000$ nodes were evaluated during indexing. Solutions up to the $1000$-th closest candidates were computed for each $N_g$. Although reporting a single solution, like the one with the lowest image difference would be sufficient, we inspected such a large number of solutions to explore where the indexing fails and understand this systematically.

\paragraph{Al-Li-Mg specimens}
The Al-Li-Mg-Ag dataset contained \SI{45.17e6}{} ions of which a fraction of \SI{4.96}{} \% were ranged as lithium and \SI{4.04}{} \% as magnesium, respectively. Amplitude spectra to build signatures for the aluminium crystal structure candidate were probed on the bin interval corresponding to $\pm$\SIrange{1.823}{2.223}{\angstrom} lattice spacing. Amplitude spectra to build signatures for the \althreeli{} crystal structure candidate were probed on the bin interval corresponding to $\pm$\SIrange{1.805}{2.206}{\angstrom} lattice spacing. This corresponds to indexing based on the \hkl<200> plane stacks of the respective sub-lattices. The same indexing settings as for the aluminium bicrystal dataset were set but now compared to three instead of one reference pattern: one reference defined by a pure aluminium crystal and two for the aluminium and the lithium sub-lattices of a \althreeli{} single crystal.

\section{Additional results}
\subsection{Robustness of the real space method against randomly missing atoms}
\begin{figure}[!htb]
\centering
\includegraphics[width=0.5\textwidth]{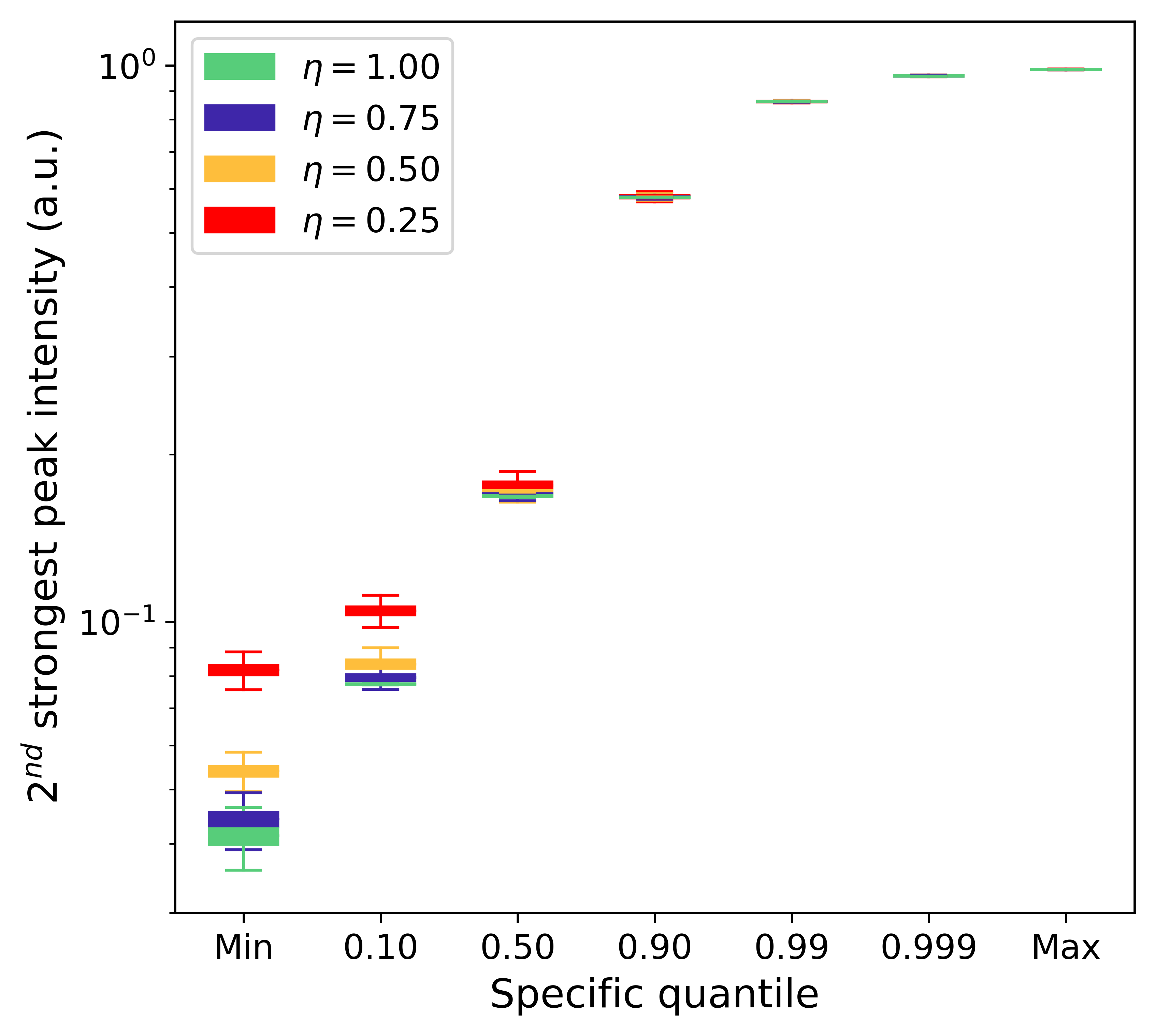}
\caption{Quantification of the real space method signal against noise due to missing atoms. For each ROI, we identified the peaks in all amplitude spectra and report the individually second strongest peak per amplitude spectrum. With $N_v$ FE mesh nodes, i.e. $N_v$ directions, this yields one cumulative distribution per ROI. Next, specific quantiles of the cumulative distribution are extracted for each ROI and displayed for the entire ROI ensemble. This condenses the statistics of how all amplitude spectra ($40962$ per ROI) for all ROIs ($10000$ in total) differ. We repeat this statistical analysis for all \ensuremath{\eta} values (the fraction of atoms remaining). The results show that the signal-to-noise ratio is not substantially affected by removing atoms.}
\label{fig:verify_noise_vacancy}
\end{figure}

Figure \ref{fig:verify_noise_vacancy} reports how the signatures computed from the perfect single crystals change when removing atoms randomly from the lattice. The results show descriptive spatial statistics for all ROIs and SDMs. The very low spread of the quantile values documents that the signatures of every ROI have peaks of similar intensity ($\approx 1.0$). The background is in all cases at least ten times lower in intensity. The observation that the spread is similar for all $\eta$ values substantiates a robustness against a random removal of atoms.

\subsection{Influence of the ROI radius and frequency resolution}
\begin{figure}[!htb]
\centering
\includegraphics[width=1.0\textwidth]{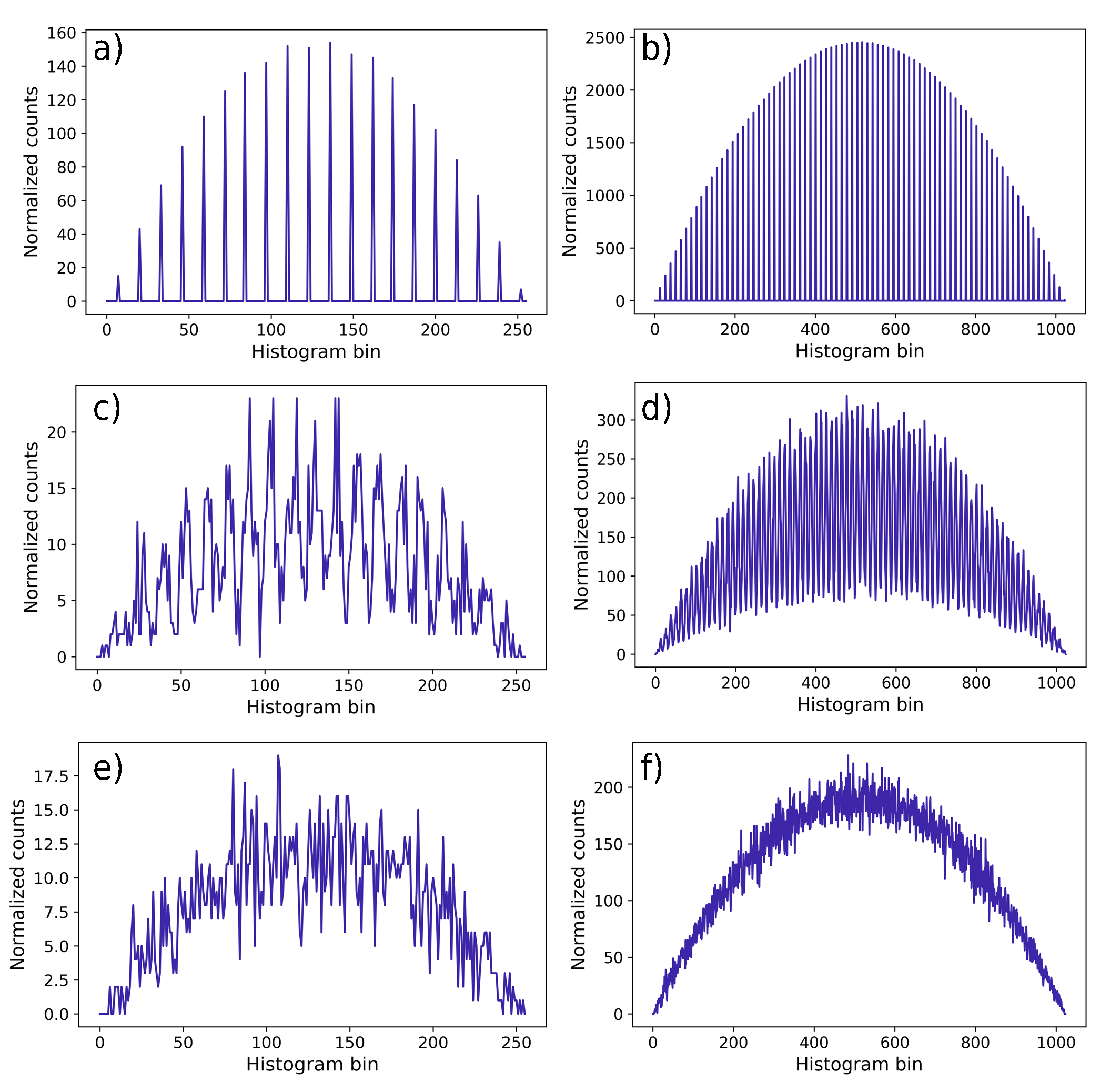}
\caption{Effect of up-scaling the radius of the ROIs for a fixed frequency resolution \ensuremath{\frac{2^{m-1}-1}{R} = \SI{6.35}{1\per\angstrom}} and increasingly stronger positional noise. The left column a), c), e) displays exemplar SDMs for \ensuremath{R = \SI{20}{\angstrom}}. The right column b), d), f) displays SDMs for the same projection direction in a), c), e) but a larger ROI radius \ensuremath{R = \SI{80}{\angstrom}}, respectively. Up-scaling the ROI increases the significance of the histogram but does not better concentrate the peaks.}
\label{fig:verify_sxindexing_roiupscaling_one}
\end{figure}

\begin{figure}[!htb]
\centering
\includegraphics[width=0.5\textwidth]{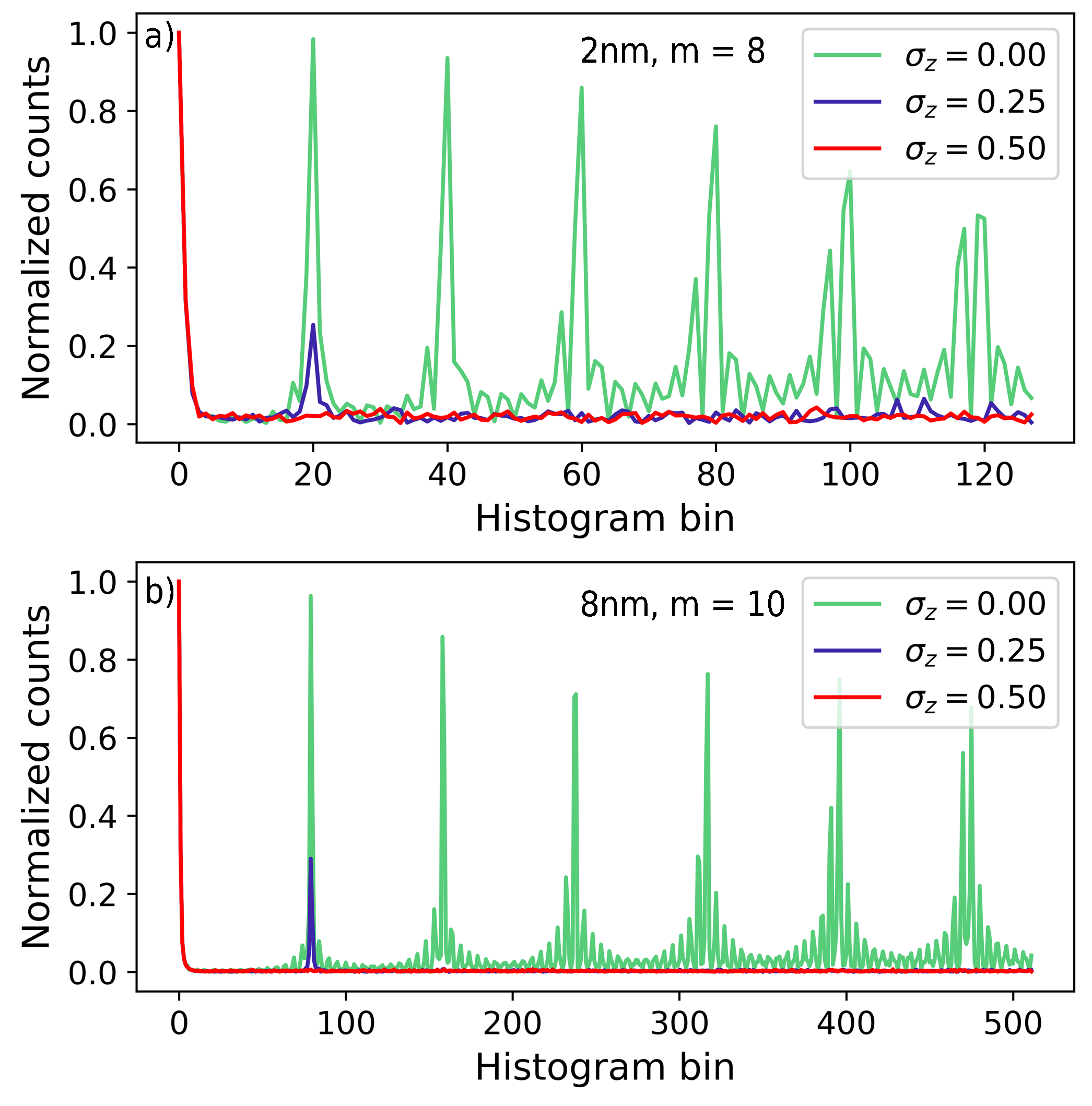}
\caption{The amplitude spectra for the above histograms support that an up-scaling of the ROI increases only the significance of the SDMs but does not better concentrate the peaks.}
\label{fig:verify_sxindexing_roiupscaling_two}
\end{figure}

Verification analyses on the same synthetic datasets but using larger radii for the ROIs did not improve the indexing quality for the noisy datasets. In fact, the exemplar SDM in Fig. \ref{fig:verify_sxindexing_roiupscaling_two} and corresponding amplitude spectra in Fig. \ref{fig:verify_sxindexing_roiupscaling_one} summarise that up-scaling the ROI only increases the total number of counts in the SDMs, i.e. the significance improves but the peaks get not better concentrated.

\newpage
\subsection{Statistical analysis of the signatures for the aluminium bicrystal}
\begin{figure}[!htb]
\centering
\includegraphics[width=0.9\textwidth]{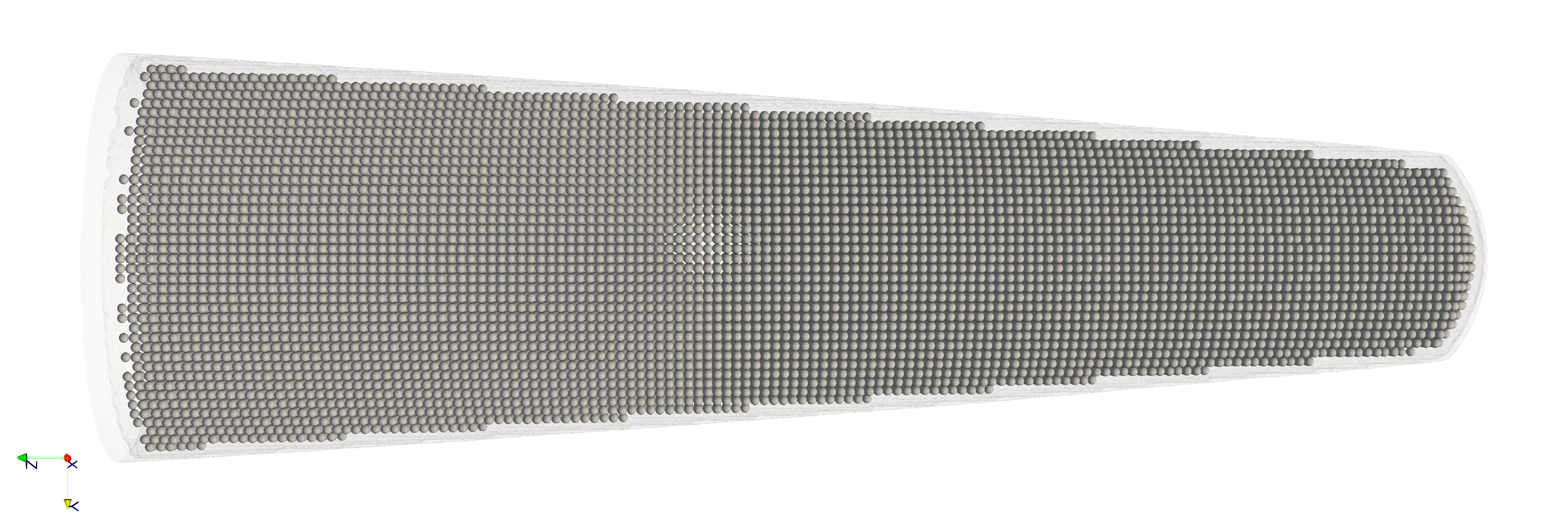}
\includegraphics[width=0.9\textwidth]{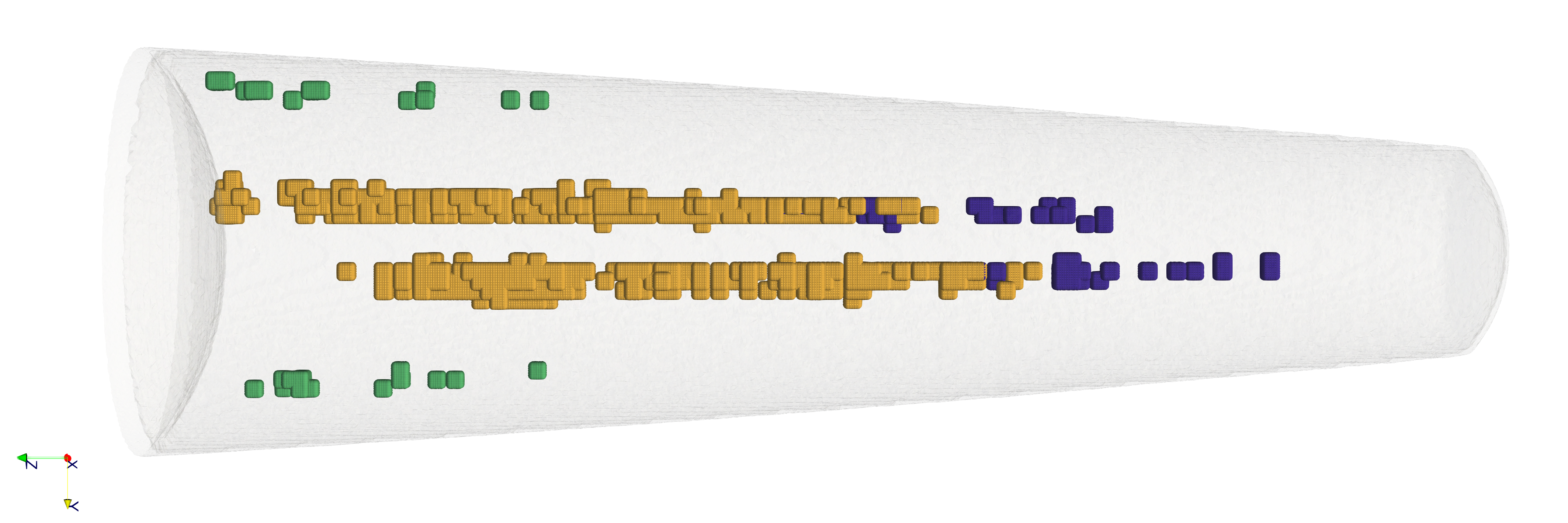}
\caption{The local ROI grid refinement allows to explore specifically those regions in the dataset more efficiently where the chance for indexing are best. The upper figure displays a coarse regular ROI grid (see explanations below). The lower figure displays signature-specific results for the environment of the refined ROIs. Different colours distinguish different signatures (\hkl{002} in dark blue, \hkl{220} in green, and \hkl{111} in orange, respectively).}
\label{fig:verify_albicrystal_grids}
\end{figure}

\begin{figure}[!htb]
\centering
\includegraphics[width=0.47\textwidth]{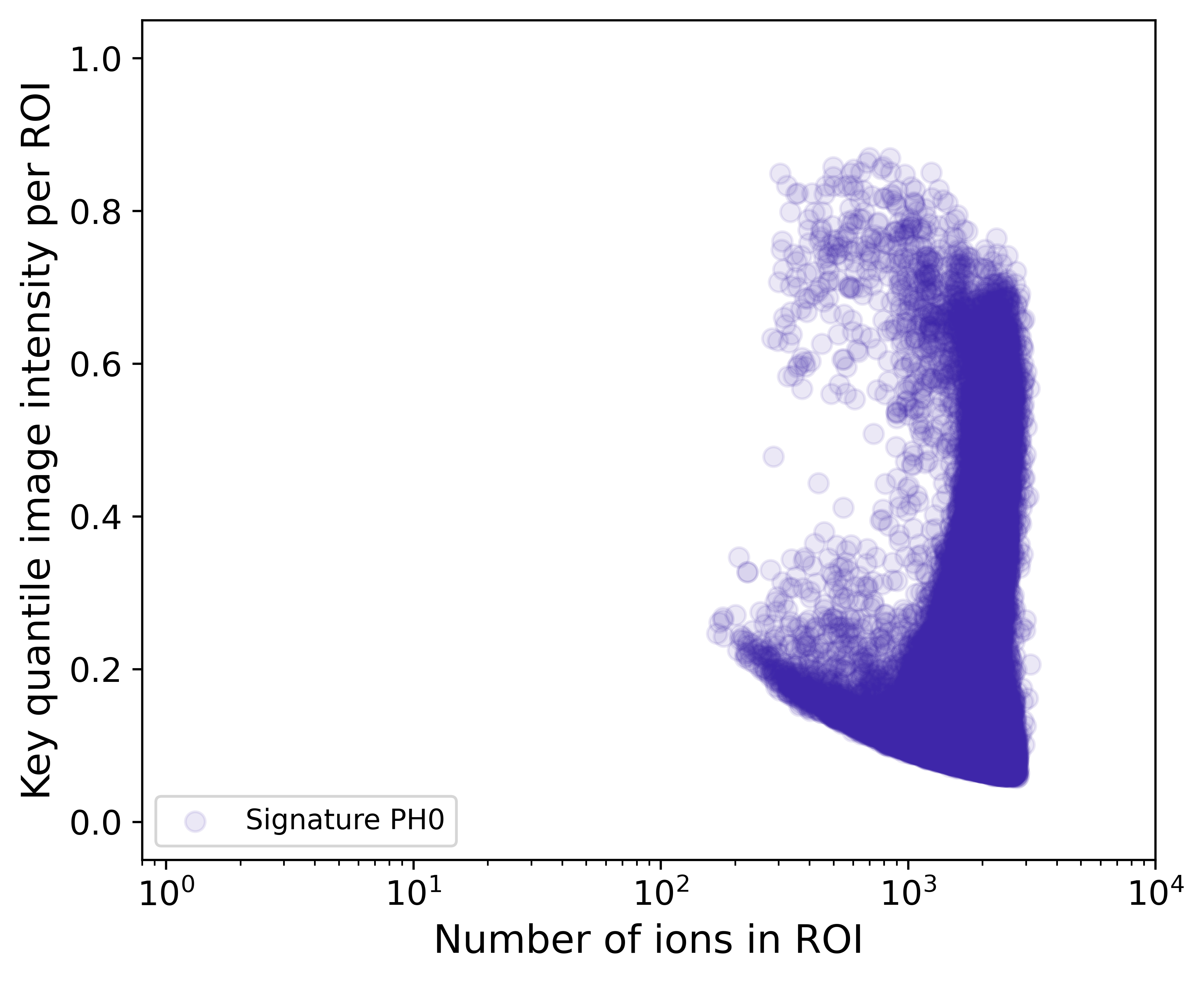}
\quad
\includegraphics[width=0.47\textwidth]{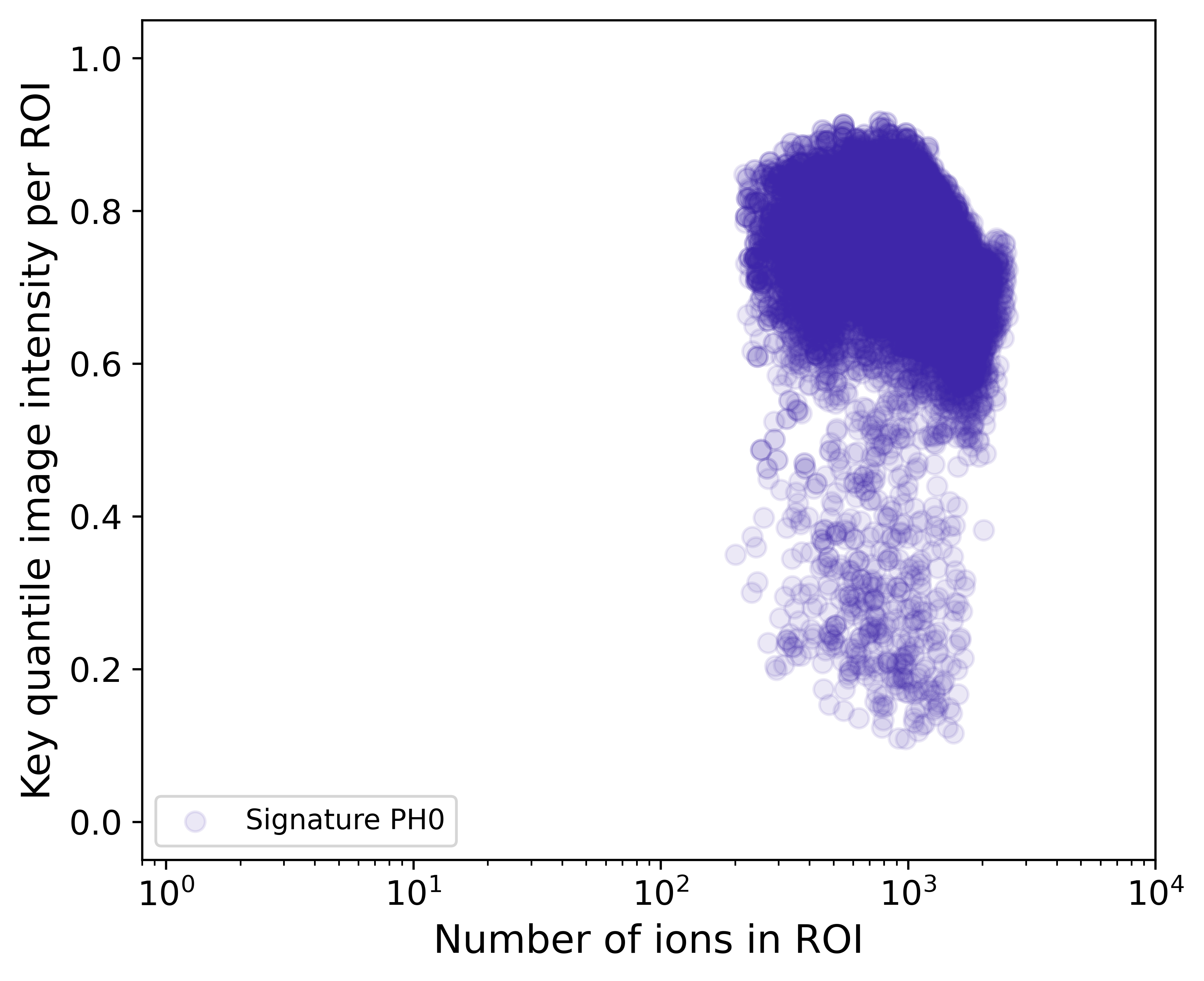}
\quad
\includegraphics[width=0.47\textwidth]{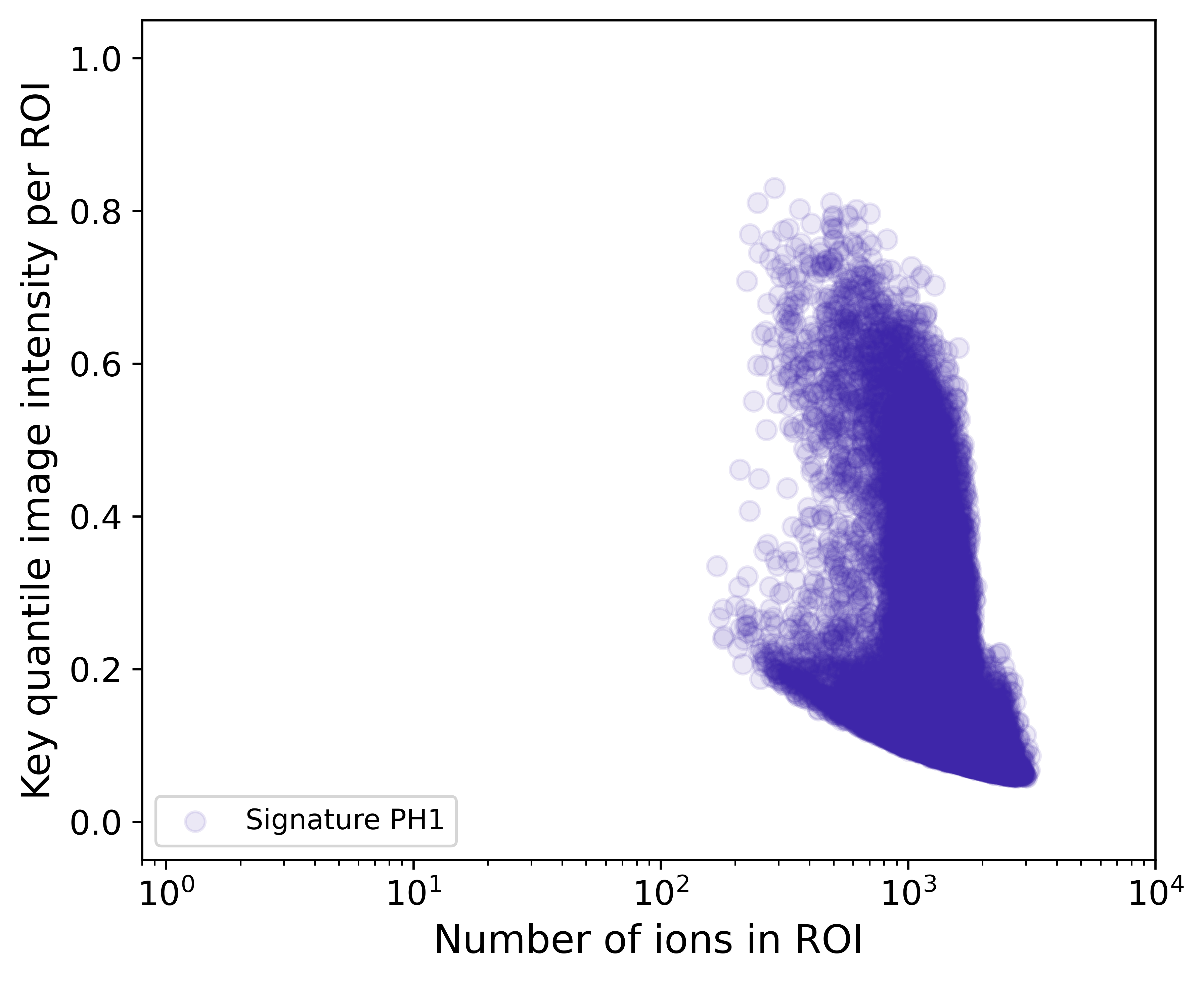}
\quad
\includegraphics[width=0.47\textwidth]{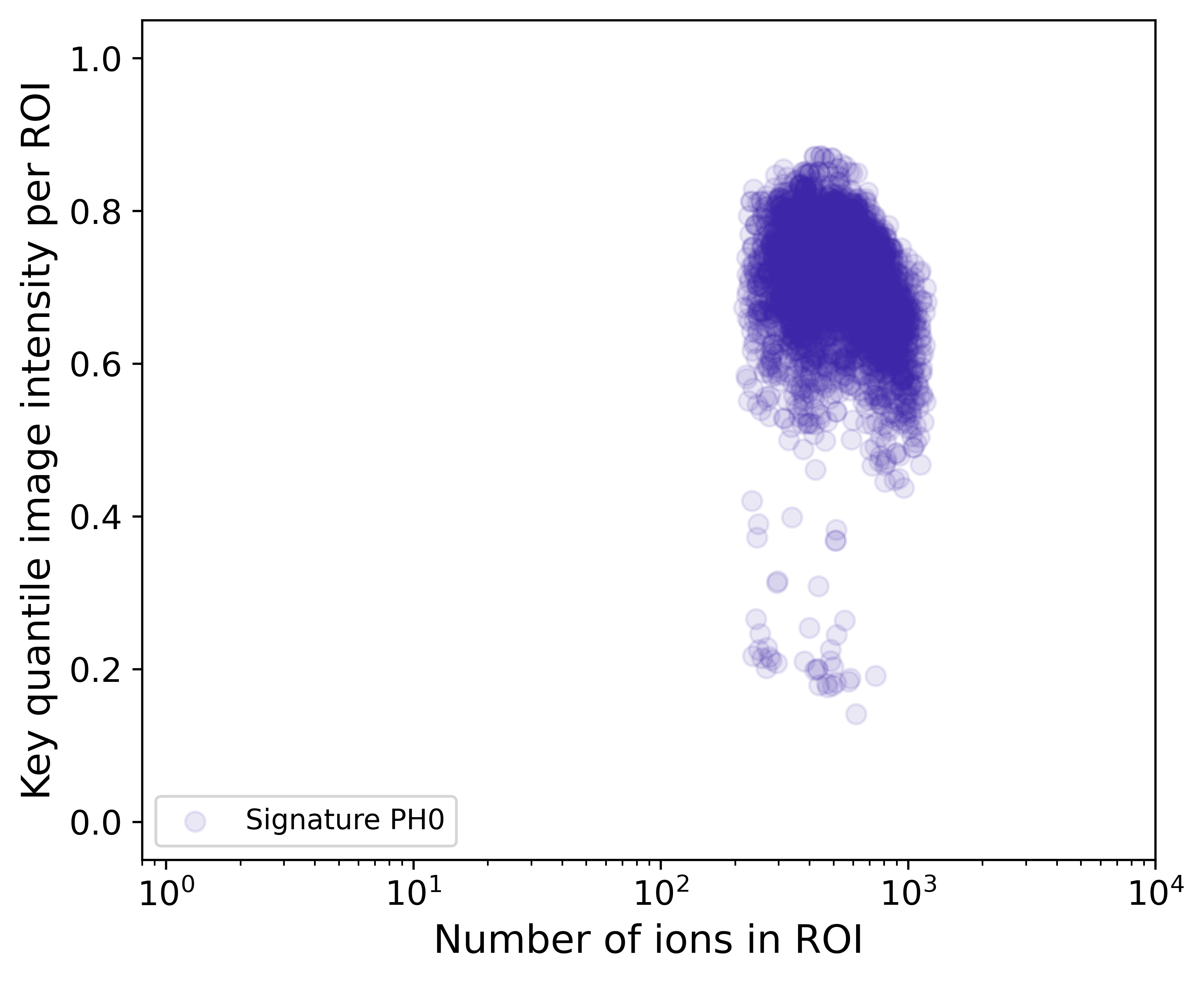}
\quad
\includegraphics[width=0.47\textwidth]{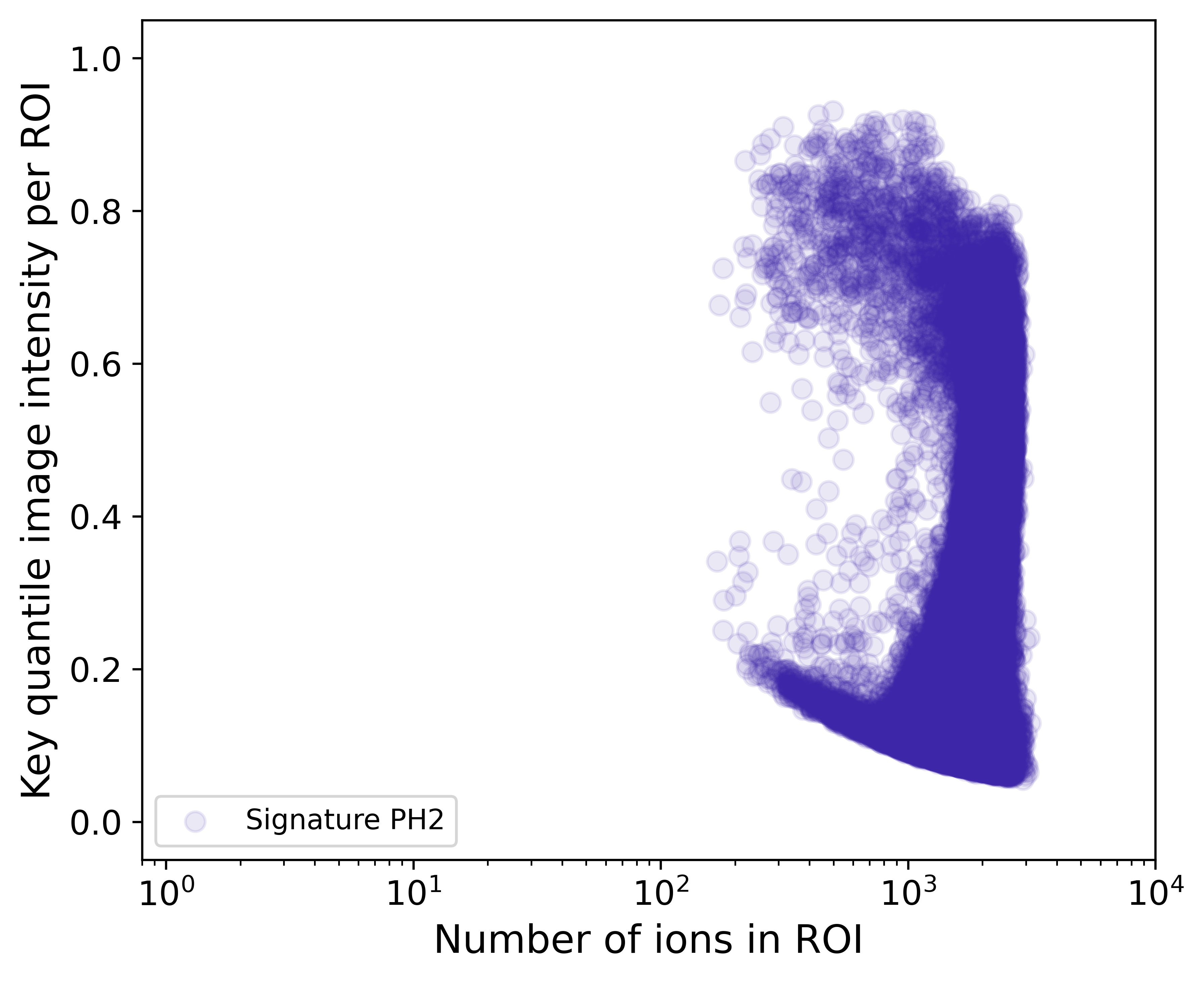}
\quad
\includegraphics[width=0.47\textwidth]{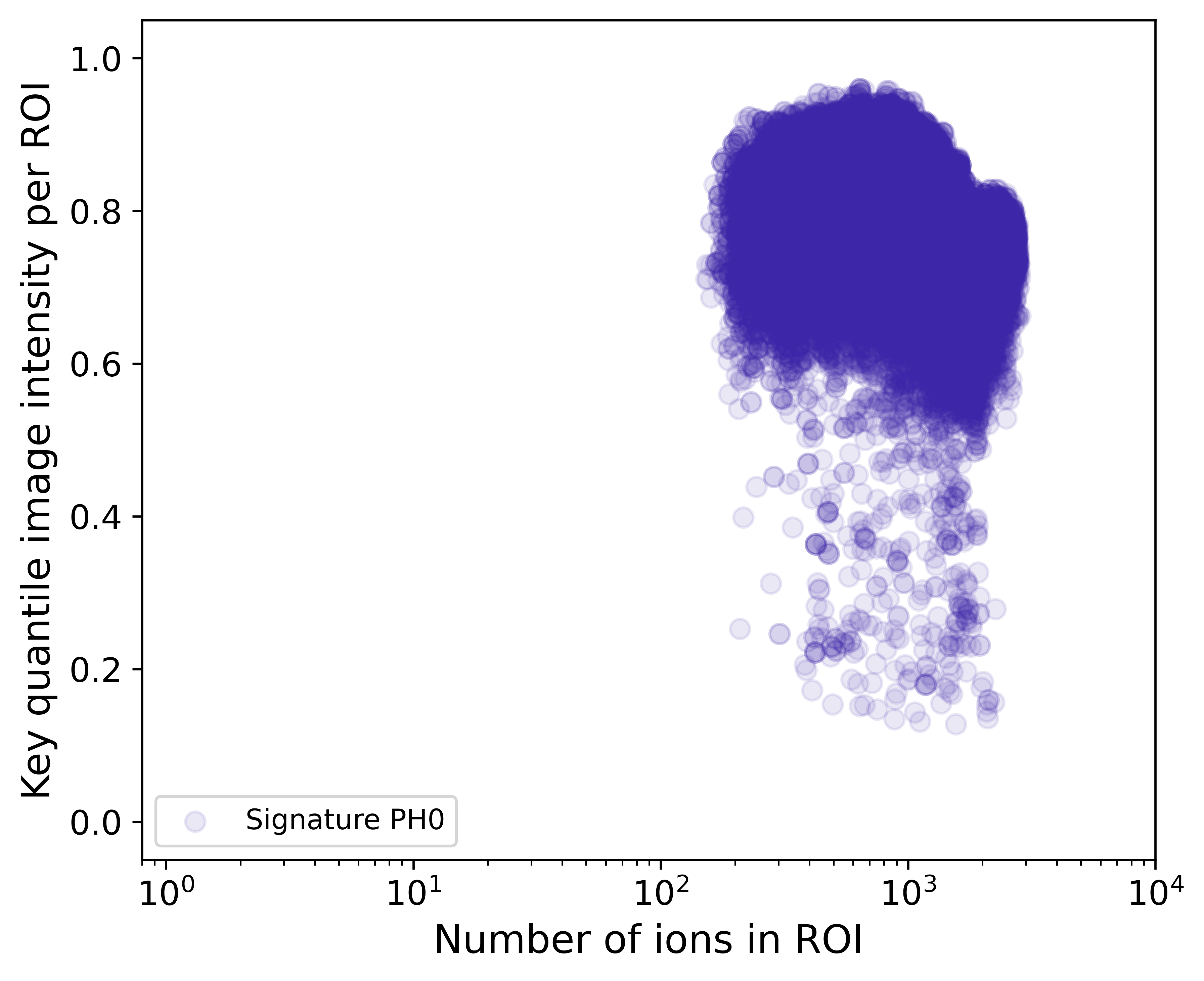}
\caption{Statistical analysis of the signal strength for three different poles \hkl{002} (upper row), \hkl{220} (middle row), and \hkl{111} (lower row), respectively. We performed an analysis with an adaptive grid and compare how many atoms are included in a ROI and how high the signal intensities are for each ROI. Specifically, we compare results for two ROI grids - a coarse one, or initial grid, respectively (left column) and a fine one, first refinement, respectively (right column). These results supplement and substantiate the findings to the aluminium bicrystal case study in the main paper. There are several regions in the dataset with strong crystallographic signal.}
\label{fig:verify_albicrystal_coarse_fine}
\end{figure}

We summarise in the main paper that it is possible to detect regions in the datasets with significant crystallographic signal. Figure \ref{fig:verify_albicrystal_grids} documents the results of a successful protocol for performing a local refinement of the ROI grid. Figure \ref{fig:verify_albicrystal_coarse_fine} supports these results with a statistical analysis of the signal strength and the number of atoms per ROI. 

Such a local grid refinement works for example as follows: First, we first scanned the dataset at a coarse resolution, here using ${\SI{20}{\angstrom}}^3$. Approximately \SI{1.0e5} ROIs were obtained as Fig. \ref{fig:verify_albicrystal_grids}a) displays (with the ROIs rendered to scale as grey spheres). Intensities were evaluated individually for each ROI. Next, we filtered out, for each pole (\hkl{hkl}) separately, those ROIs whose maximum image intensity was at least $0.75$. Finally, we performed a grid refinement for these ROIs (${\SI{0.5}{\nano\meter}}^3$). In effect, between \SIrange{4.0e3}{8.3e4}{} now more closely spaced ROIs were characterised. The dataset volume in Fig. \ref{fig:verify_albicrystal_grids}b) shows that these ROIs probe different regions of the dataset. By contrast, a naive grid refinement of the coarse grid would have resulted in $64$ times more computations, so our local refinement is at least one order of magnitude more efficient.

\newpage
\subsection{Statistical analysis of the signatures for the Al-Li-Mg-Ag dataset}
\begin{figure}[!htb]
\centering
\includegraphics[width=0.47\textwidth]{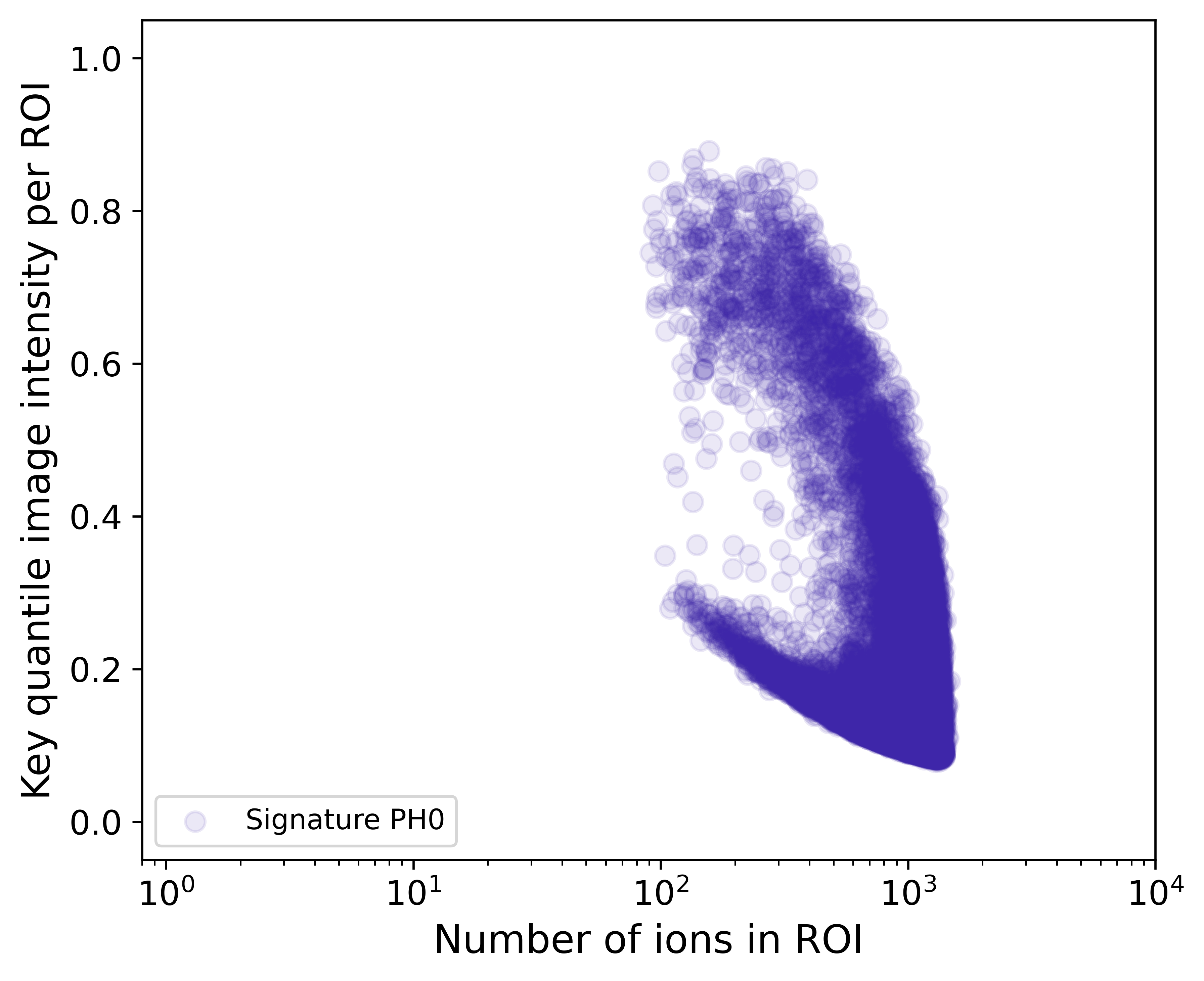}
\quad
\includegraphics[width=0.47\textwidth]{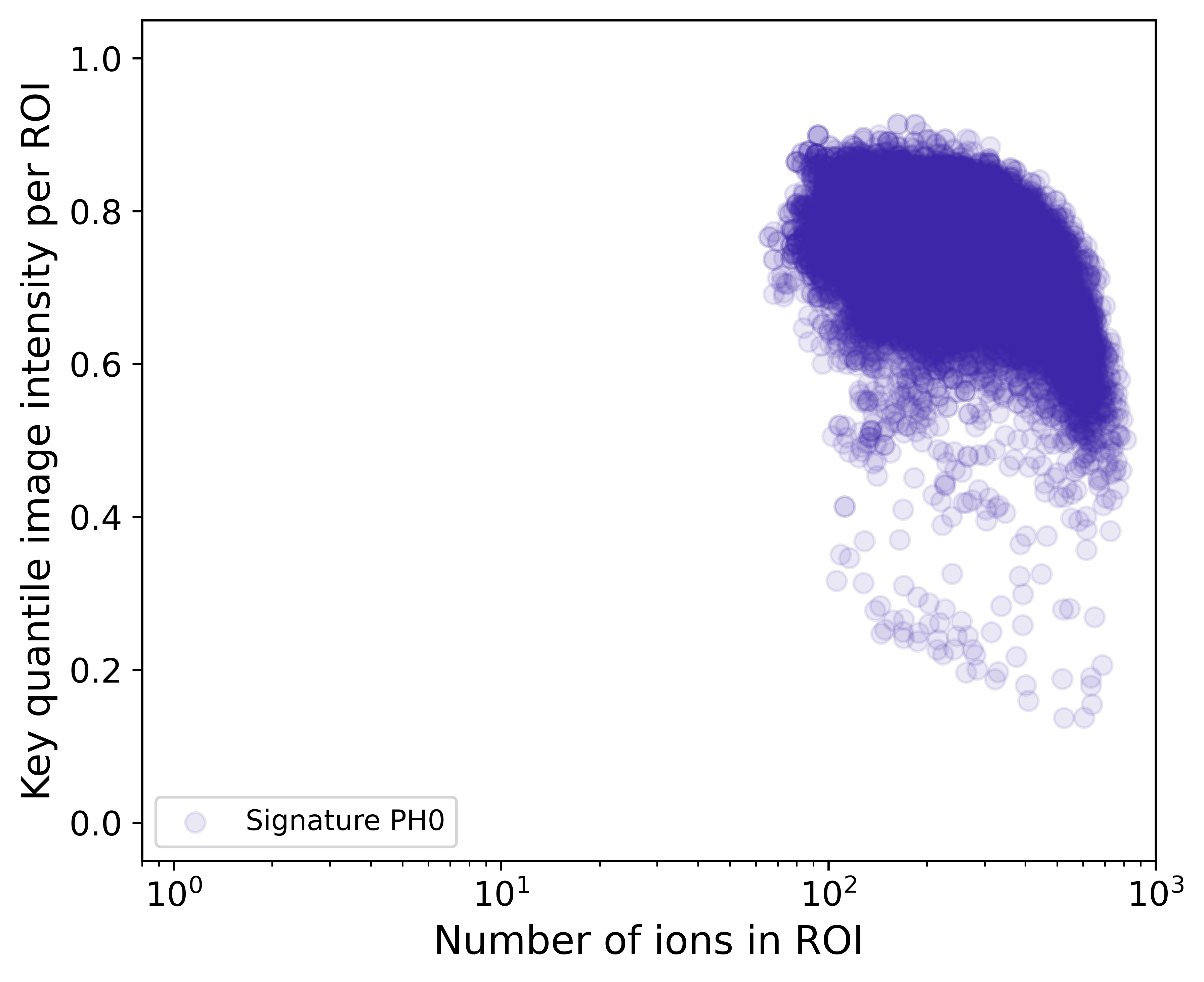}
\quad
\includegraphics[width=0.47\textwidth]{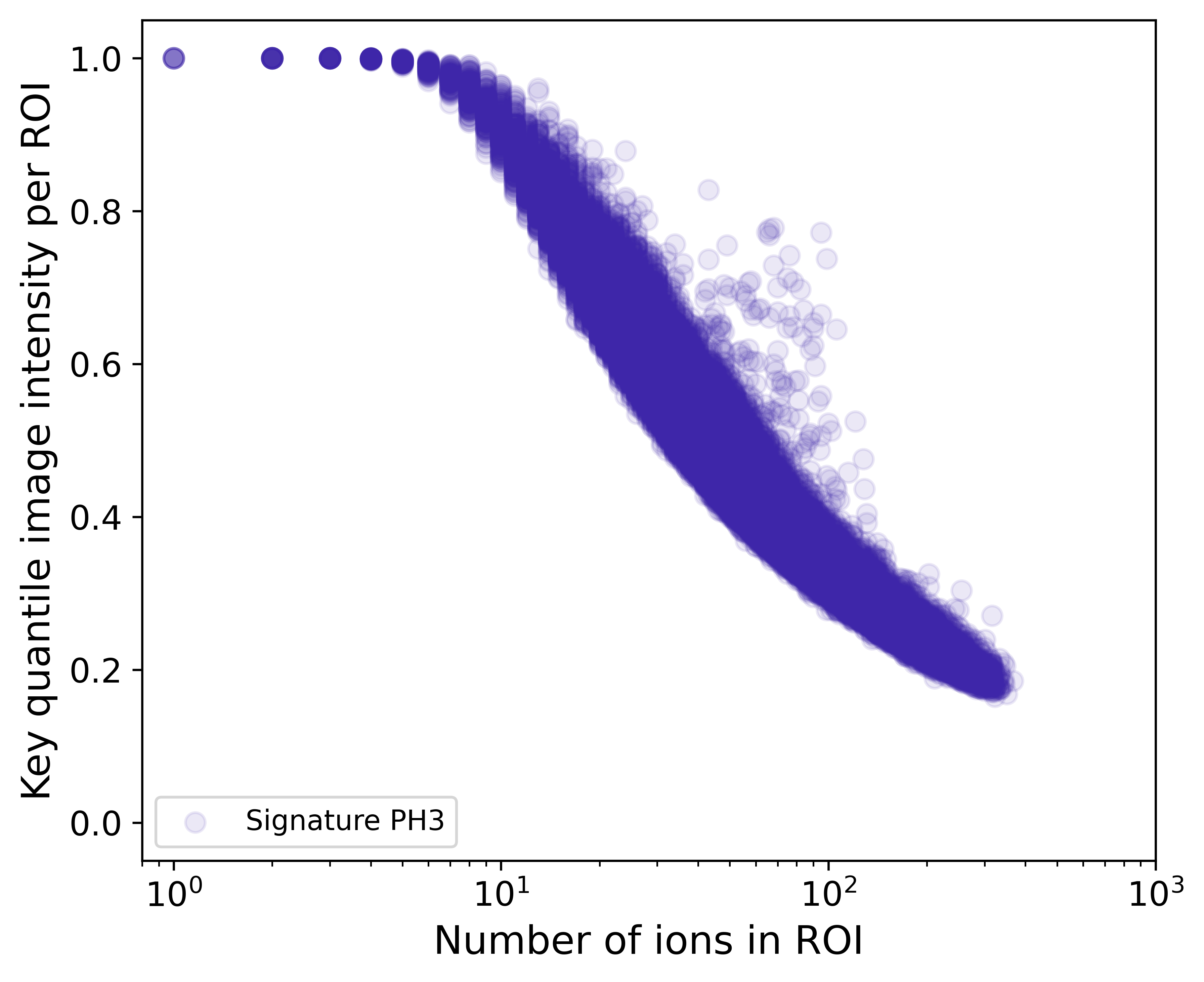}
\caption{Statistical analysis of the signal strength for \hkl{002} signatures. The upper row shows statistics for aluminium atoms and compares the coarse (results to the left) with the fine grid (results to the right, using the same thresholding of image intensities like in Fig. \ref{fig:verify_albicrystal_coarse_fine}). Below is the result for the coarse grid and \hkl{002} signatures computed based on the lithium atoms. For further methodological details see text and explanation of the figure above. The results back-up the conclusion of the main paper that finite counting effects can result in strong peaks and have to be carefully separated from ROIs with more significant signal. The lower the number of atoms in the ROI the noisier and eventually even skewed the SDMs become. When fast-Fourier-transforming these SDMs, this translates into high peaks in the amplitude spectra, i.e. high image intensities in the signatures. This is especially pronounced for ROIs which contain less than a few dozen lithium atoms.}
\label{fig:verify_allimg_coarse_fine}
\end{figure}

We performed the above statistical analyses with grid refinement also for the Al-Li-Mg-Ag dataset. The same poles as the ones above were analysed but now two signatures computed for each \hkl{hkl} - one signature for the aluminium and one for the lithium sub-lattice. Figure \ref{fig:verify_allimg_coarse_fine} shows the results, exemplified for \hkl{002} signatures.

\newpage
\subsection{Benchmarking: reasons for limited scalability}
There are multiple reasons for limited scalability: $10000$ ROIs cannot be equally distributed on $3200$ CPUs. In effect, work load differences limit the scalability. In addition, there are a few code portions that remained sequentially executed. A detailed analysis of the elapsed time data showed that there are costs of synchronisation bottlenecks. These could be worth a future inspection and software optimisation. However, reducing these bottlenecks will require efforts and implementation techniques that are beyond the interest and toolkit of most atom probe practitioners and should therefore not be pursued here.

\subsection{Preliminary analysis of compressing signatures with spherical harmonics}
In what follows, we summarise the mathematical details of the discussed spherical harmonics compressing. First, we discretised a unit sphere into a geodesic sphere finite element mesh to allow for a mesh of near equal-area elements and a near constant density of nodal points per unit surface area. The spherical image (the signatures) to be fit is evaluated at the nodal points, resulting in a vector of values, $\left\{ \epsilon \right\}$ (length $n^n$, the number of nodes on the geodesic mesh). A set of spherical harmonic modes are similarly defined at the nodal points of the geodesic mesh, given as $\left\{ H^k \right\}$ (length $n^n$). With the spherical harmonic modes defined, the spherical image is described via a series expansion:
\begin{equation}
    \left\{ \epsilon \right\} = \sum_{k=1}^{n^h}{w^k \left\{ H^k \right\}},
\end{equation}
where $n^h$ is the number of harmonic modes in the series expansion. To solve for the weights, $w^k$, a method of least squares is used:
\begin{equation}
    \left[H\right]^T\left[H\right] \left\{w\right\} = \left[H\right]^T \left\{ \epsilon \right\}
\end{equation}
Here, $\left\{w\right\}$ is a vector of the weights of the series expansion (length $n^h$), and $\left[H\right]$ is a matrix of dimension $n^n \times n^h$, constructed of nodal values of each harmonic mode:
\begin{equation}
    \left[ H \right] = \left[ \left\{ H^1 \right\} \left\{ H^2 \right\} \left\{ H^3 \right\} ... \left\{ H^{n^h} \right\} \right]
\end{equation}

\begin{figure}[!htb]
\centering
\includegraphics[width=0.5\textwidth]{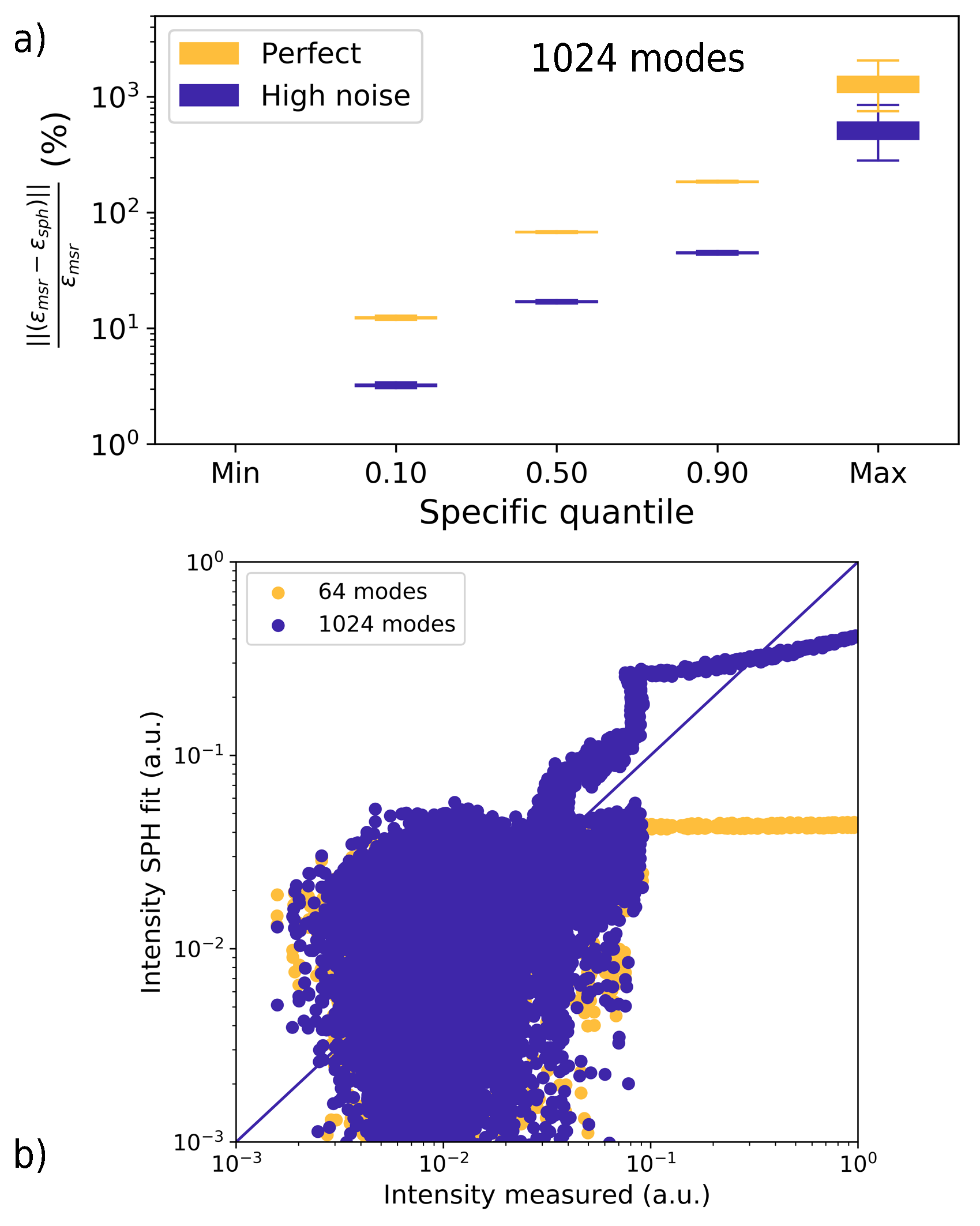}
\caption{Approximating signatures of the aluminium single crystal synthetic datasets with a series of spherical harmonics. Peak positions are in principle captured but the intensities remain smeared out even for a very large number of modes.}
\label{fig:matthew_sph_detail}
\end{figure}

The resulting vector of harmonic weights yields a reduced-order description of the spherical image. We discuss in the main paper that the spherical harmonics approach captures the location of the peaks. However, it smears out the intensity substantially. The QQ plot in Fig. \ref{fig:matthew_sph_detail} quantifies this visual impression.

\newpage
\printbibliography

\end{document}